\documentclass[preprint]{ptephy_v1}
\usepackage{amssymb}
\usepackage{graphicx}
\numberwithin{equation}{section}

\newcommand{\cb}{\bar{c}}
\newcommand{\lb}[1]{\label{#1}}
\newcommand{\cl}{\mathcal{L}}
\newcommand{\bb}{\bar{B}}

\newcommand{\vp}{\varphi}
\newcommand{\ptl}{\partial}
\newcommand{\ve}{\varepsilon}

\newcommand{\bt}{\tilde{B}}

\newcommand{\mcb}{\mathcal{B}}
\def\bve#1{\mbox{\boldmath $#1$}}
\def\mc#1{\mathcal{#1}}

\begin{document}

\title{Condensates, massive gauge fields and confinement in 
the SU(3) gauge theory
}


\author{Hirohumi Sawayanagi\thanks{National Institute of Technology, Kushiro College, Kushiro, 084-0916, Japan\\\quad E-mail:sawa@kushiro-ct.ac.jp}}




\begin{abstract}
     SU(3) gauge theory in the nonlinear gauge of the Curci--Ferrari type is studied.  In the low-energy region, 
ghost condensation and subsequent gauge field condensation can happen.  The latter condensation makes 
classical gauge fields massive.  If the color electric potential with string is chosen as the classical gauge 
field, it produces the static potential with the linear potential.  We apply this static potential to 
the three-quark system, and show, different from the $Y$-type potential, infrared divergence remains in the 
$\Delta$-type potential.  The color electric flux is also studied, and show 
that the current which plays the role of the magnetic current appears.  
\end{abstract}

\subjectindex{B0,B3,B6}

\maketitle

\section{Introduction}

     In the dual superconductor picture of quark confinement, it is expected that monopole condensation appears 
in the low energy region.  This condensation produces a mass of gauge fields, and confinement happens.    
To describe this scenario, the dual Ginzburg--Landau model has been considered (See, e.g., Ref.~\cite{rip}).     

     Based on the SU(2) gauge theory in a nonlinear gauge, we considered another possibility to give a mass for gauge 
fields \cite{hs17}.  In the low-energy region 
below $\Lambda_{\mathrm{QCD}}$, which is the QCD scale parameter, the ghost condensation happens.  Although this condensation 
gives rise to a tachyonic gluon mass, a gauge field condensate $\langle A^{+\mu}A^-_{\mu}\rangle$ can remove the tachyonic 
mass.  If there is a classical U(1) gauge field, this classical field becomes massive by this condensate.  

     In Ref.~\cite{hs19}, referring to the Zwanziger's formalism \cite{zwa}, the electric potential and its dual potential were 
introduced as a classical field.  Due to the string structure of these classical fields, the linear potential was obtained.  

     In this paper, we extend the previous approach to the SU(3) case.       
In the next section, the ghost condensation is studied at the one-loop level.  In Sect. 3, under the ghost condensation, 
tachyonic gluon masses and the gluon condensates 
$\langle A^{a\mu}A^a_{\mu}\rangle$ are calculated in the low momentum limit.  The Lagrangian of the massive classical gauge fields 
is also presented.  In Sect. 4, as the classical fields, the color electric potential and its dual potential are introduced, 
and the static potential between two charges is calculated.  
Using the result of Sect. 4, the mesonic potential and the baryonic potential are discussed in Sect. 5.  
Different from the dual Ginzburg--Landau model, there is no magnetic current originally.  
The Maxwell's equations in the present model are studied in Sect. 6.  The behavior of the color flux tube is also 
considered.  
Section 7 is devoted to summary and comment.  In Appendix A, the relation between the ghost condensation and 
$\Lambda_{\mathrm{QCD}}$ is derived.  Tachyonic gluon masses are calculated in Appendix B.  
To make the article self-contained, an example of the electric potential and its dual potential for 
a color charge is presented in Appendix C.  In Appendix D, the static potential between two charges is calculated in detail.   
Based on a phenomenological Lagrangian for order parameters, the type of the dual superconductivity in the present model 
is considered in Appendix E.

\section{Ghost condensation }

\subsection{Notation}

       We consider the SU(3) gauge theory with structure constants $f_{abc}$ in the Minkowski space.  
The Lagrangian in the nonlinear gauge of the Curci--Ferrari type \cite{cf} is given by \cite{hs03}
\begin{align*}
 \cl &= \cl_{inv}+ \cl_{NL},\quad \cl_{inv}=-\frac{1}{4}F_{\mu\nu}^aF^{a\mu\nu}, \notag \\
  \cl_{NL} &=  B^a \partial_{\mu}A^{a\mu}+i\cb^a(\partial_{\mu}D^{\mu}c)^a
 + \frac{\alpha_1}{2}B^aB^a + \frac{\alpha_2}{2}\bb^a\bb^a -B^aw^a, \quad (a=1,\cdots,8)  
\end{align*}
where $B^a$ is the Nakanishi--Lautrup field, $c$ $(\cb)$ is the ghost (antighost), 
$\bb^a = -B^a+ ig f_{abc}\cb^b c^c$, $\alpha_1$ and $\alpha_2$ are gauge parameters, and 
$w^a$ is a constant to keep the BRS symmetry.  The Lagrangian $\cl_{NL}$ is rewritten as 
\begin{equation*}
 \cl_{\vp}=\frac{\alpha_1}{2}B^aB^a 
 +B^a (\ptl_{\mu}A^{a\mu}+\vp^a -w^a)+i\cb^a \left[(\ptl_{\mu}D^{\mu})^{ac}+gf_{abc}\vp^b\right]c^c 
 -\frac{\vp^a\vp^a}{2\alpha_2},  
\end{equation*}
where the auxiliary field $\varphi^a$ represents $-\alpha_2 \bb^a$.

          Let us expand the gauge field $A_{\mu}$ as 
\begin{equation}
 A_{\mu}=A^a_{\mu}\frac{\lambda^a}{2}=\vec{A}_{\mu}\cdot \vec{H} + \sum_{\alpha=1}^3 \left( W_{\mu}^{-{\alpha}}E_{\alpha}
+  W_{\mu}^{\alpha}E_{-\alpha}\right), \lb{201}
\end{equation}
where the diagonal components are
\[ \vec{A}_{\mu}=(A_{\mu}^3, A_{\mu}^8),\quad \vec{H}=(H^3, H^8)=\left(\frac{\lambda^3}{2}, \frac{\lambda^8}{2}\right),  \]
and 
$\vec{A}_{\mu}\cdot \vec{H}=\sum_{A=3,8}A^A_{\mu} H^A =A^3_{\mu} H^3+ A^8_{\mu} H^8$.  The off-diagonal components are given by  

\[ W_{\mu}^{\pm 1}= \frac{1}{\sqrt{2}}\left(A_{\mu}^1 \pm iA_{\mu}^2\right), \quad W_{\mu}^{\pm 2}= \frac{1}{\sqrt{2}}\left(A_{\mu}^4 \mp iA_{\mu}^5\right), 
\quad W_{\mu}^{\pm 3}= \frac{1}{\sqrt{2}}\left(A_{\mu}^6 \pm iA_{\mu}^7\right), \]
\[ E_{\pm 1}=\frac{1}{2\sqrt{2}}(\lambda^1 \pm i\lambda^2), \quad
 E_{\pm 2}=\frac{1}{2\sqrt{2}}(\lambda^4 \mp i\lambda^5), \quad E_{\pm 3}=\frac{1}{2\sqrt{2}}(\lambda^6 \pm i\lambda^7).  
\]
Using the root vectors of the SU(3) group  
\begin{equation}
 \vec{\epsilon}_1=(1,0),\quad  \vec{\epsilon}_2=\left(\frac{-1}{2},\frac{-\sqrt{3}}{2}\right),\quad
 \vec{\epsilon}_3=\left(\frac{-1}{2},\frac{\sqrt{3}}{2}\right), \lb{202}
\end{equation}
the above matrices satisfy 
\[ [\vec{H}, E_{\pm \alpha}]=\pm \vec{\epsilon}_{\alpha} E_{\pm \alpha},\quad 
[E_{\alpha}, E_{-\alpha}]=\vec{\epsilon}_{\alpha} \cdot \vec{H}, \quad
[E_{\pm \alpha}, E_{\pm \beta}]=\frac{\mp 1}{\sqrt{2}}\ve_{\alpha \beta \gamma}E_{\mp \gamma} \]
and 
\[ \mathrm{tr}\left(H^AH^B\right)=\frac{\delta^{AB}}{2},\quad \mathrm{tr}\left(E_{\alpha}E_{-\beta}\right)=\frac{\delta^{\alpha\beta}}{2}.  
\]

     In the same way, $c$ and $\cb$ are expressed as 
\begin{equation}
c=\vec{c}\cdot\vec{H} + \sum_{\alpha=1}^3 \left( C^{-\alpha}E_{\alpha}
+  C^{\alpha}E_{-\alpha}\right),  \quad
\bar{c} =\vec{\bar{c}}\cdot \vec{H} + \sum_{\alpha=1}^3 \left( \bar{C}^{-\alpha}E_{\alpha}
+  \bar{C}^{\alpha}E_{-\alpha}\right), \lb{203}    
\end{equation}
where 
\[ C^{\pm 1}= \frac{1}{\sqrt{2}}\left(c^1 \pm i c^2\right), \quad C^{\pm 2}= \frac{1}{\sqrt{2}}\left(c^4 \mp i c^5\right), 
\quad C^{\pm 3}= \frac{1}{\sqrt{2}}\left(c^6 \pm i c^7\right),  \]
and $\bar{C}^{\pm \alpha}\ (\alpha=1,2,3) $ are defined as well.

\subsection{Ghost condensation}

     To obtain the one-loop effective potential of $\vp^A$, we diagonalize $\displaystyle \varphi^a\frac{\lambda^a}{2}$ as 
$\vec{\vp}\cdot\vec{H}$.  
Then, using the expressions (\ref{201}) and (\ref{203}), the Lagrangian 
$i\bar{c}^a\Box c^a +  i\bar{c}^a gf_{abc}\vp^b c^c
=2\mathrm{tr}(i\bar{c}\Box c + \bar{c}[g\vp,c]) $ becomes 
\begin{equation}   
  \sum_{A=3,8}\bar{c}^A i \Box c^A +
 \sum_{\alpha=1}^3\left\{\bar{C}^{\alpha}(i\Box+g\vec{\epsilon}_{\alpha}\cdot\vec{\vp}) C^{-\alpha}
 + \bar{C}^{-\alpha}(i\Box-g\vec{\epsilon}_{\alpha}\cdot\vec{\vp}) C^{\alpha} \right\}.   \lb{204}
\end{equation}

     Next, as in the SU(2) case \cite{hs07}, we integrate out $C^{\pm \alpha}$ and $\bar{C}^{\mp \alpha}$ with the 
momentum $\mu \leq k \leq \Lambda$.  After the Wick rotation, we obtain the potential 
\begin{align*}
   \sum_{\alpha=1}^3 V_1\left(\vec{\epsilon}_{\alpha}\cdot\vec{\varphi}\right) &= 
-\sum_{\alpha=1}^3\int_{\mu}^{\Lambda} \frac{d^4k}{(2\pi)^4} \ln \left[(-k^2)^2 +g^2(\vec{\epsilon}_{\alpha}\cdot\vec{\varphi})^2 \right]\\
 &=-\frac{1}{32\pi^2}\sum_{\alpha=1}^3\left[ \left\{\Lambda^4+g^2(\vec{\epsilon}_{\alpha}\cdot\vec{\varphi})^2 \right\}\ln\left\{\Lambda^4+g^2(\vec{\epsilon}_{\alpha}\cdot\vec{\varphi})^2 \right\}  \right.\\
 &\hspace{1.8cm} \left.-\left\{\mu^4+g^2(\vec{\epsilon}_{\alpha}\cdot\vec{\varphi})^2 \right\}\ln\left\{\mu^4+g^2(\vec{\epsilon}_{\alpha}\cdot\vec{\varphi})^2 \right\} \right]. 
\end{align*}
Since we can rewrite $\vp^a\vp^a/(2\alpha_2)$ as  
\[
\frac{1}{2\alpha_2}\vp^a\vp^a=\frac{1}{2\alpha_2}\vec{\vp}\cdot\vec{\vp}=\frac{1}{3\alpha_2}\sum_{\alpha=1}^3 (\vec{\epsilon}_{\alpha}\cdot\vec{\varphi})^2, 
\]
the one-loop effective potential of $\varphi$ becomes \cite{ks} 
\begin{equation}
  V(\varphi)=
\sum_{\alpha=1}^3 \left[\frac{(\vec{\epsilon}_{\alpha}\cdot\vec{\vp})^2}{3\alpha_2} 
 + V_1\left(\vec{\epsilon}_{\alpha}\cdot\vec{\vp}\right) \right].  \lb{205}
\end{equation}

     To study minimum points of $V(\vp)$, we consider 
\begin{align*}
 \frac{\ptl V}{\ptl \vp^8}&=\frac{\vp^8}{\alpha_2}-\frac{3}{2}g^2\vp^8\left\{L(\vec{\epsilon}_{2}\cdot\vec{\vp})+L(\vec{\epsilon}_{3}\cdot\vec{\vp})\right\}
-\frac{\sqrt{3}}{2}g^2\vp^3\left\{L(\vec{\epsilon}_{2}\cdot\vec{\vp})-L(\vec{\epsilon}_{3}\cdot\vec{\vp})\right\} , \\
 \frac{\ptl V}{\ptl \vp^3}&=\frac{\vp^3}{\alpha_2}-g^2\vp^3\left[2L(\vec{\epsilon}_{1}\cdot\vec{\vp})
+\frac{1}{2}\left\{ L(\vec{\epsilon}_{2}\cdot\vec{\vp})+L(\vec{\epsilon}_{3}\cdot\vec{\vp})\right\}\right]
-\frac{\sqrt{3}}{2}g^2\vp^8\left\{L(\vec{\epsilon}_{2}\cdot\vec{\vp})-L(\vec{\epsilon}_{3}\cdot\vec{\vp})\right\},  \\ 
&L(\vec{\epsilon}_{\alpha}\cdot\vec{\vp})=\frac{1}{32\pi^2}\ln \left\{ \frac{\Lambda^4+(g\vec{\epsilon}_{\alpha}\cdot\vec{\vp})^2}{\mu^4+(g\vec{\epsilon}_{\alpha}\cdot\vec{\vp})^2}\right\}. 
\end{align*}
The explicit forms of $g\vec{\epsilon}_{\alpha}\cdot\vec{\vp}\ (\alpha=1,2,3)$ are 
\[ g\vec{\epsilon}_{1}\cdot\vec{\varphi}=g\vp^3 ,\quad g\vec{\epsilon}_{2}\cdot\vec{\varphi}=-\frac{g}{2}\left(\vp^3+ \sqrt{3}\vp^8\right),\quad 
g\vec{\epsilon}_{3}\cdot\vec{\varphi}=\frac{g}{2}\left(-\vp^3+\sqrt{3}\vp^8\right),  \]
and $\vp^8=0$ 
leads to $\vec{\epsilon}_{2}\cdot \vec{\varphi}=\vec{\epsilon}_{3}\cdot \vec{\varphi}$.  
So, the equation 
$\ptl V/\ptl \vp^8=0$ has the solution $\vp^8=0$.  Now we assume 
$\langle g\vec{\vp}\rangle=(v, 0)$ is a minimum point.  Since the potential $V(\vp)$ is invariant under the interchange 
$\vec{\epsilon}_{\alpha}\cdot\vec{\vp} \longleftrightarrow \vec{\epsilon}_{\beta}\cdot\vec{\vp}\ (\alpha \neq \beta)$, and has the symmetry 
$\vec{\epsilon}_{\alpha}\cdot\vec{\vp} \to -\vec{\epsilon}_{\alpha}\cdot\vec{\vp}$,   
there are six minimum points 
\footnote{These minimum points were found in Ref.~\cite{ks}.  It also 
contains the three-dimensional figure of $V(\vp)$.  }  

\begin{align}
 &(v,0),\quad \left(-\frac{v}{2},-\frac{\sqrt{3}}{2}v\right),\quad \left(-\frac{v}{2},\frac{\sqrt{3}}{2}v\right), \nonumber \\
 &(-v,0),\quad \left(\frac{v}{2},\frac{\sqrt{3}}{2}v\right),\quad \left(\frac{v}{2},-\frac{\sqrt{3}}{2}v\right).  \lb{206}  
\end{align}

     To determine the value of $v$, we consider the case $(v,0)$. 
The condition $\ptl V/\ptl \vp^3=0$ with $g\vp^3=v\neq 0$ becomes 
\begin{equation}
 \frac{32\pi^2}{\alpha_2g^2}=\ln\left\{\left(\frac{v^2+\Lambda^4}{v^2+\mu^4}\right)^2\left(\frac{v^2+4\Lambda^4}{v^2+4\mu^4}\right)\right\}.  \lb{207}
\end{equation}
If we set $v=0$ at $\mu=\mu_0$, $\mu_0=\Lambda \exp[-8\pi^2/3\alpha_2g^2]$ is obtained.  
When the cut-off $\Lambda$ is large enough,  Eq.(\ref{207}) gives 
$v\simeq 2^{1/3}\mu_0^2$ in the limit $\mu \to 0$.

       In Appendix~A, we show $3\alpha_2=\beta_0$  is the ultraviolet fixed point of $\alpha_2$, where 
$\beta_0=11N/3$ with $N=3$ is the first coefficient of the $\beta$ function.    
Substituting this value into $\mu_0$, we find 
\[ \mu_0=\Lambda \exp\left[-\frac{8\pi^2}{\beta_0g^2}\right]=\Lambda_{\mathrm{QCD}}, \] 
where $\Lambda_{\mathrm{QCD}}$ is the QCD scale parameter.  
Thus we obtain the ghost condensate $v$ that behaves as  
\begin{equation*}
 v =0 \  (\mu \geq \Lambda_{\mathrm{QCD}}), \quad v\neq 0 \  (\mu<\Lambda_{\mathrm{QCD}}), 
\quad v \simeq 2^{1/3}\Lambda_{\mathrm{QCD}}^2 \  (\mu\to 0).  
\end{equation*}

\section{Gluon mass}

\subsection{Tachyonic gluon mass}

     In the SU(2) gauge theory, ghost loops with $v\neq 0$ produce the tachyonic gluon mass terms \cite{hs03, dv}.  
To study the SU(3) case, we choose the vacuum $(v, 0)$ in Eq.(\ref{206}), and write $g\vp^a=v\delta^{3a}+g\tilde{\vp}^a$, where $\tilde{\vp}^a$ 
is the quantum part.  
Neglecting $\tilde{\vp}^a$, the Lagrangian (\ref{204}) becomes 
\begin{equation*}
\sum_{A=3,8}i\bar{c}^A\Box c^A 
+ \sum_{\alpha=1}^3\left\{ i\bar{C}^{\alpha}(\Box -i\epsilon^3_{\alpha}v)C^{-\alpha} +i\bar{C}^{-\alpha}(\Box +i\epsilon^3_{\alpha}v)C^{\alpha} \right\}, 
\end{equation*}
and it leads to the ghost propagators  
\begin{align}
 & \langle c^A\bar{c}^A \rangle =  -\frac{i}{\Box}\quad (A=3,8), \nonumber \\ 
 & \langle C^{\alpha}\bar{C}^{-\alpha} \rangle = -\frac{i}{\Box +i\epsilon^3_{\alpha}v},\quad  
  \langle C^{-\alpha}\bar{C}^{\alpha} \rangle = -\frac{i}{\Box -i\epsilon^3_{\alpha}v}\quad (\alpha=1,2,3).  \lb{301}
\end{align}

    Using Eqs.(\ref{201}) and (\ref{203}), the vertex $-i\ptl_{\mu}\bar{c}^agf_{abc}A^{b\mu}c^c$ in $i\cb^a (\ptl_{\mu}D^{\mu})^{ab}c^b$ is rewritten as 
\begin{align}
 &\sum_{A=3,8}\left[-gA^{A\mu}\sum_{\alpha=1}^3 \epsilon_{\alpha}^A\left\{ (\ptl_{\mu}\bar{C}^{\alpha)})C^{-\alpha}
- (\ptl_{\mu}\bar{C}^{-\alpha})C^{\alpha} \right\}
+g(\ptl_{\mu}\bar{c}^A)\sum_{\alpha=1}^3\epsilon_{\alpha}^A(W^{\alpha\mu}C^{-\alpha}-W^{-\alpha\mu}C^{\alpha}) \right.  
\nonumber \\
&\left.-g\sum_{\alpha=1}^3\epsilon_{\alpha}^A\left\{W^{\alpha\mu}(\ptl_{\mu}\bar{C}^{-\alpha})-
W^{-\alpha\mu}(\ptl_{\mu}\bar{C}^{\alpha})\right\}c^A \right]
+\sum_{(\alpha,\beta,\gamma)}\mathrm{sgn}(\gamma)\frac{g}{\sqrt{2}}\epsilon_{\alpha\beta\gamma}
(\ptl_{\mu}\bar{C}^{\alpha})C^{\beta}W^{\gamma \mu},
\lb{302}
\end{align}
where $\mathrm{sgn}(\gamma)$ is the sign of $\gamma$, and $\sum_{(\alpha,\beta,\gamma)}$ implies the sum for 
the permutations of  
$(1,2,3)$ and $(-1,-2,-3)$.

     Now we consider the two-point function 
\begin{equation}
   \langle A^{a}_{\mu}(x)A^b_{\nu}(y)\rangle =G^{ab}_{\mu\nu}(x,y)  \lb{303}
\end{equation}
at the one-loop level.  As in the SU(2) case, using the ghost propagators (\ref{301}) and the interactions (\ref{302}), 
the ghost loop in Fig.1(b) gives rise to tachyonic masses in the low momentum 
limit $p\to 0$.  The details are presented in Appendix~B.  
From Eqs.(\ref{b03}) and (\ref{b07}), we find the tachyonic mass terms are
\begin{equation}
 -\frac{1}{2}\left(\frac{5}{2}m^2\right)A^{3\mu}A_{\mu}^3 -\frac{1}{2}\left(\frac{3}{2}m^2\right)A^{8\mu}A_{\mu}^8 
-\sum_{\alpha=1}^3\left(\frac{5}{4}m^2\right)W^{\alpha\mu}W^{-\alpha}_{\mu}, \quad m^2=\frac{g^2 v}{64\pi}.  
\lb{304}
\end{equation}

\begin{figure}
\begin{center}
\includegraphics{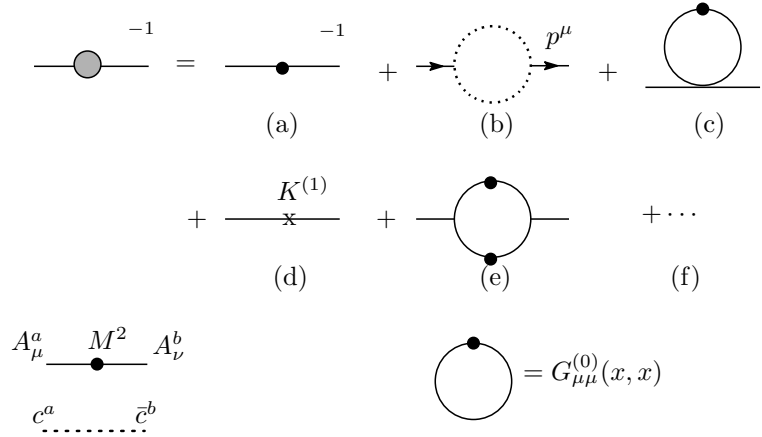}
\caption{The diagrams that contribute to the inverse propagator for $A_{\mu}^{a}$.  Fig.~1(b) gives rise to the tachyonic mass in the 
limit $p \to 0$, and Fig.~1(c) yields the condensate $G_{\mu\mu}^{(0)}(x,x)$.}
\label{fig1}
\end{center}
\end{figure}

\subsection{Condensate $\langle A^{a\mu}A^a_{\mu}\rangle$}

     To remove the tachyonic masses, we consider the condensate $\langle A^{a\mu}A^a_{\mu}\rangle$ \cite{hs17,hs19}.  
Let us introduce the source terms 
\[ 
     \sum_{A=3,8}K_AA^{A\mu}A^A_{\mu} + \sum_{\alpha=1}^3 \mathcal{K}_{\alpha} W^{\alpha\mu}W^{\alpha}_{\mu}.   
\]
Although the sources may depend on the momentum scale, for simplicity, the constant sources 
\begin{align*} 
  K_A&=K_A^{(0)}+K_A^{(1)}+\cdots,\quad K_A^{(n)}=O(\hslash^n), \quad K_A^{(0)}=\frac{1}{2}M_A^2   \quad (A=3,8)
\\
   \mc{K}_{\alpha}&=\mc{K}_{\alpha}^{(0)}+\mc{K}_{\alpha}^{(1)}+\cdots, \quad \mc{K}_{\alpha}^{(n)}=O(\hslash^n), \quad
\mc{K}_{\alpha}^{(0)}=\mc{M}_{\alpha}^2 \quad (\alpha=1,2,3)  
\end{align*}
are considered.  The interaction $-(gf_{abc}A^b_{\mu}A^c_{\nu})^2/4$ in $-(F^a_{\mu\nu})^2/4$ contains the terms 
\begin{align}
&-\frac{g^2}{2}\sum_{\alpha=1}^3 W^{\alpha\mu}W^{-\alpha}_{\mu}W^{\alpha\nu}W^{-\alpha}_{\nu}
-\frac{g^2}{4}\sum_{\alpha\neq \beta}W^{\alpha\mu}W^{-\alpha}_{\mu}W^{\beta\nu}W^{-\beta}_{\nu} \nonumber \\
&-g^2\sum_{\alpha=1}^3
\left(\vec{\epsilon}_{\alpha}\cdot\vec{A}^{\mu}\right)\left(\vec{\epsilon}_{\alpha}\cdot\vec{A}_{\mu}\right)W^{\alpha\nu}W^{-\alpha}_{\nu}.  \lb{305}
\end{align}
So, at $O(\hslash)$, the diagram in Fig.1(c)  
gives the condensate 
$\langle A^{a\mu}A^a_{\mu}\rangle^{(0)}=g^{\mu\nu}G^{(0)aa}_{\mu\nu}(x,x)$, where $G^{(0)ab}_{\mu\nu}(x,y)$ is the free propagator 
with the mass $M_A$ or $\mc{M}_{\alpha}$.  If the other divergent diagrams of $O(\hslash)$ are subtracted by the terms with 
$K_{A}^{(1)}$ or $\mc{K}_{\alpha}^{(1)}$, the condensate 
$\langle A^{a\mu}A^a_{\mu}\rangle^{(0)}$ is determined to remove the tachyonic masses in Eq.(\ref{304}).

     As an example, we consider the self-energy of $W^{1}_{\mu}$ in the limit $p\to 0$.  
The diagram Fig.1(c) with the first interaction in Eq.(\ref{305}) gives 
$-g^2\langle W^{1\mu}W^{-1}_{\mu}\rangle^{(0)}$.  Similarly, from the second term in Eq.(\ref{305}), we obtain 
$-g^2\left[\langle W^{2\mu}W^{-2}_{\mu}\rangle^{(0)}+\langle W^{3\mu}W^{-3}_{\mu}\rangle^{(0)}\right]/2$.  
Since $\vec{\epsilon}_1\cdot \vec{A}_{\mu}=A^{3}_{\mu}$, the third term in Eq.(\ref{305}) gives 
$-g^2\langle A^{3\mu}A^3_{\mu}\rangle^{(0)}$.  
So, Fig.1(c) for  $W^1_{\mu}$ leads to   
\[  -g^2\left\{\mc{G}^1+\frac{1}{2}\left(\mc{G}^2+\mc{G}^3\right)+G^3\right\},   \]
where $\mc{G}^{\alpha}=\langle W^{\alpha\mu}W^{-\alpha}_{\mu}\rangle^{(0)}\ (\alpha=1,2,3)$ and 
$G^A=\langle A^{A\mu}A^A_{\mu}\rangle^{(0)}\ (A=3,8)$.   
The condition that these condensates remove the tachyonic mass of $W^1_{\mu}$ becomes 
\begin{equation*}
   -\frac{5}{4}m^2-g^2\left\{\mc{G}^1+\frac{1}{2}\left(\mc{G}^2+\mc{G}^3\right)+G^3\right\}=0.  
\end{equation*}  

     In the same way, we find the conditions   
\begin{align}
& -\frac{5}{4}m^2-g^2\left\{\mc{G}^2+\frac{1}{2}\left(\mc{G}^3+\mc{G}^1\right)+\frac{1}{4}\left(G^3+3G^8\right)\right\}=0, 
 \lb{306} \\ 
& -\frac{5}{4}m^2-g^2\left\{\mc{G}^3+\frac{1}{2}\left(\mc{G}^1+\mc{G}^2\right)+\frac{1}{4}\left(G^3+3G^8\right)\right\}=0, 
\nonumber \\  
& -\frac{5}{4}m^2-g^2\left\{\mc{G}^1+\frac{1}{4}\left(\mc{G}^2+\mc{G}^3\right)\right\}=0, \lb{307} \\ 
& -\frac{3}{4}m^2-\frac{3g^2}{4}\left(\mc{G}^2+\mc{G}^3\right)=0, \lb{308}  
\end{align}
for $W^{2}_{\mu}$, $W^{3}_{\mu}$, $A^{3}_{\mu}$ and $A^{8}_{\mu}$, respectively.  

     The solutions of these five equations are 
\begin{equation}
  \mc{G}^1=-\frac{m^2}{g^2},\quad \mc{G}^{2}=\mc{G}^{3}=-\frac{m^2}{2g^2},\quad G^3=\frac{m^2}{4g^2},\quad 
G^8=-\frac{m^2}{12g^2}.  \lb{309}  
\end{equation}
We note, although the diagonal component $\langle A^{3\mu}A^3_{\mu}\rangle^{(0)}$ vanishes in the SU(2) case \cite{hs17, hs19}, 
the diagonal components $\langle A^{A\mu}A^{A}_{\mu}\rangle^{(0)}\ (A=3,8)$ do not vanish in SU(3).

\subsection{Inclusion of classical solutions}

     To incorporate U(1)$_3$ and U(1)$_8$ classical solutions into the above scheme, we divide $A^A_{\mu}$ into the classical part 
$b^A_{\mu}$ and the quantum fluctuation $a^A_{\mu}$ as 
\[  A^A_{\mu} = b^A_{\mu} + a^A_{\mu}  \quad  (A=3,8),   \]
and divide the gauge transformation $\delta A_{\mu}=D_{\mu}(A)\varepsilon$ as 
\begin{equation}
   \delta a_{\mu}=D_{\mu}(a,W)\varepsilon,\quad (\delta b_{\mu})^a=g f_{abc}b^b\varepsilon^c,\quad 
\delta W_{\mu}=D_{\mu}(a,W)\varepsilon,  \lb{310}
\end{equation}
where $D_{\mu}(A)^{ab}=\ptl_{\mu}\delta^{ab}+g f_{acb}A_{\mu}^c$, and $D_{\mu}(a,W)$ is obtained by removing 
$b_{\mu}$ from $D_{\mu}(A)$, i.e., 
$D_{\mu}(a,W)=D_{\mu}(A)|_{b_{\mu}=0}$.  Using the gauge-fixing function 
$G(a,W)=\ptl_{\mu}A^{\mu}|_{b_{\nu}=0}+\vp -w$, the transformation (\ref{310}) gives the ghost Lagrangian 
\[
\left. i\cb^a \left[\left(\ptl_{\mu}D^{\mu}(A)\right)^{ac}+gf_{abc}\vp^b\right]c^c\right|_{b_{\mu}=0}=
i\cb^a \left[\left(\ptl_{\mu}D^{\mu}(a,W)\right)^{ac}+gf_{abc}\vp^b\right]c^c. 
\]
So, after the ghost condensation, the tachyonic mass terms are obtained by replacing $A^a_{\mu}$ with 
$a^A_{\mu}$ and $W^{\alpha}_{\mu}$ as 
\footnote{We can use the background covariant gauge.  In this case, as the ghost Lagrangian is 
$i\cb^a \left[\left(D(b)_{\mu}D^{\mu}(A)\right)^{ac}+gf_{abc}\vp^b\right]c^c$, $\cb$ and $c$ couple with $b_{\mu}$.  
However, this ghost Lagrangian has the $\mathrm{U(1)}_3\times \mathrm{U(1)}_8$ 
symmetry $\delta_{\ve}b_{\mu}=-\ptl_{\mu}\vec{\ve}\cdot \vec{H}/g$, $\delta_{\ve}a_{\mu}=0$ and 
$\delta_{\ve}W^{\pm \alpha}_{\mu}=\mp i \vec{\ve}\cdot \vec{\epsilon}_{\alpha}W^{\pm \alpha}_{\mu}$.  
Therefore, as in the SU(2) case \cite{hs17}, 
this symmetry prevents $b_{\mu}^A$ from getting tachyonic mass terms, and Eq.(\ref{311}) is obtained.  }
\begin{equation}
  -\frac{1}{2}\left(\frac{5}{2}m^2\right)a^{3\mu}a_{\mu}^3 -\frac{1}{2}\left(\frac{3}{2}m^2\right)a^{8\mu}a_{\mu}^8 
-\sum_{\alpha=1}^3\left(\frac{5}{4}m^2\right)W^{\alpha\mu}W^{-\alpha}_{\mu}.  \lb{311}
\end{equation}

     The above tachyonic mass terms are removed by the condensates 
$\mc{G}^{\alpha}=\langle W^{\alpha\mu}W^{-\alpha}_{\mu}\rangle^{(0)}\ (\alpha=1,2,3)$ and 
$G^A=\langle a^{A\mu}a^A_{\mu}\rangle^{(0)}\ (A=3,8)$ in Eq.(\ref{309}).  
When $\mc{G}^{\alpha}\neq 0$, the interaction
\[
-g^2\sum_{\alpha=1}^3
\left(\vec{\epsilon}_{\alpha}\cdot(\vec{a}+\vec{b})^{\mu}\right)\left(\vec{\epsilon}_{\alpha}\cdot(\vec{a}+\vec{b})_{\mu}\right)W^{\alpha\nu}W^{-\alpha}_{\nu}  
\]
in Eq.(\ref{305}) leads to the mass terms 
\begin{equation*}
 -g^2\sum_{\alpha=1}^3(\vec{\epsilon}_{\alpha}\cdot\vec{b}^{\mu})(\vec{\epsilon}_{\alpha}\cdot\vec{b}_{\mu})\mc{G}^{\alpha}
=-g^2\left\{\mc{G}^1+\frac{1}{4}\left(\mc{G}^2+\mc{G}^3\right)\right\}b^{3\mu}b^3_{\mu}
-\frac{3g^2}{4}\left(\mc{G}^2+\mc{G}^3\right)b^{8\mu}b^8_{\mu}.  
\end{equation*} 
Since the classical part $b^A_{\mu}$ has no tachyonic mass, the equations (\ref{307}) and (\ref{308}) imply that 
these mass terms become  
\begin{equation} 
    \sum_{A=3,8}\frac{m_A^2}{2}b^{A\mu}b^A_{\mu}, \quad m_3^2=\frac{5m^2}{2},\quad m_8^2=\frac{3m^2}{2}.  
\lb{312}
\end{equation}

     Thus, after integrating out $c$ and $\cb$, we obtain the low-energy effective Lagrangian   
\begin{align}
   \cl_{\mathrm{eff}}=&  \cl_{\mathrm{cl}}+\sum_{A=3,8}\left\{
 -\frac{1}{4}(\ptl \wedge a^A)^{\mu\nu}(\ptl \wedge a^A)_{\mu\nu} + \frac{M_A^2}{2}a^{A\mu}a^A_{\mu} \right\} \nonumber \\
&+\sum_{\alpha=1}^3\left\{ -\frac{1}{4}(\ptl \wedge W^{\alpha})^{\mu\nu}(\ptl \wedge W^{\alpha})_{\mu\nu}
  +\mc{M}_{\alpha}^2 W^{\alpha\mu}W^{\alpha}_{\mu} \right\} +\cdots, \nonumber \\
 \cl_{\mathrm{cl}}=&\sum_{A=3,8}\left\{-\frac{1}{4}(\ptl \wedge b^A)^{\mu\nu}(\ptl \wedge b^A)_{\mu\nu} + \frac{m_A^2}{2}b^{A\mu}b^A_{\mu}
\right\},  \lb{313}
\end{align}
where $(\ptl \wedge A^a)_{\mu\nu}=\ptl_{\mu}A^a_{\nu}-\ptl_{\nu}A^a_{\mu}$.   

        We ignored the momentum dependence of the sources $K_A$ and $\mathcal{K}_{\alpha}$, and 
applied the $\hslash$ expansion.  Because it is difficult to modify this treatment, 
we use $\mathcal{L}_{\mathrm{cl}}$ as the first approximation of the low energy Lagrangian.

\section{Classical fields and static potential}

\subsection{The classical electric potential $\bt^A_{\mu}$ and its dual potential $\mcb^A_{\mu}$} 

     It is expected that the Abelian component of the gauge field dominates in confinement \cite{ei}.  
Based on the previous works \cite{hs18, hs19}, 
we choose the dual electric potential $\mcb_{\mu}^A$ as the classical field $b_{\mu}^A\ (A=3,8)$.  
It describes the electric monopole solution \cite{hs19}.  
The color electric current $j_{\mu}^A$ is incorporated by the replacement 
\[ (\ptl \wedge \mcb^A)^{\mu\nu}  \to\ ^dF^{A\mu\nu}=
 (\ptl \wedge \mcb^A)^{\mu\nu} + \epsilon^{\mu\nu\alpha\beta}(n\cdot\ptl)^{-1}n_{\alpha}j^A_{\beta},  \]
where the space-like vector $n^{\mu}$ \cite{zwa} is chosen as $n^{\mu}=(0,\bve{n})$ with $|\bve{n}|=1$, and 
$n\cdot \ptl=n^{\mu}\ptl_{\mu}$.  
We note this is the Zwanziger's dual field strength $F^d=(\ptl \wedge B)+(n \cdot \ptl)^{-1}(n \wedge j_e)^d$ in Ref.~\cite{zwa}.  
Then  the Lagrangian .(\ref{313}) becomes 
\begin{equation}
 \cl_{\mathrm{cl}}= \sum_{A=3,8} \left[-\frac{1}{4}\left\{(\ptl \wedge \mcb^A)^{\mu\nu} + 
\epsilon^{\mu\nu\alpha\beta}(n\cdot\ptl)^{-1}n_{\alpha}j^A_{\beta}\right\}^2
   + \frac{m_A^2}{2}\left(\mcb_{\mu}^A\right)^2\right].  \lb{401}
\end{equation}
The equation of motion for $\mcb_{\mu}^A$ is 
\begin{equation}
   (D_{m_A}^{-1})^{\mu\nu}\mcb_{\nu}^A= -\epsilon^{\mu\rho\alpha\beta}(n\cdot\ptl)^{-1}n_{\rho}\ptl_{\alpha}j_{\beta}^A,
\quad (D_{m_A}^{-1})^{\mu\nu}=(\square +m_A^2)g^{\mu\nu} - \ptl^{\mu}\ptl^{\nu},  \lb{402}
\end{equation} 
and $\mcb_{\mu}^A$ is solved as 
\begin{equation}
         \mcb_{\mu}^A=-(D_{m_A})_{\mu\nu}\epsilon^{\nu\rho\alpha\beta}(n\cdot\ptl)^{-1}n_{\rho}\ptl_{\alpha}j_{\beta}^A, \quad 
(D_{m_A})_{\mu\nu}=\frac{g_{\mu\nu}-\ptl_{\mu}\ptl_{\nu}/\square}{\square+m_A^2}
+\frac{\ptl_{\mu}\ptl_{\nu}}{m_A^2\square}.  \lb{403}
\end{equation}
If we use Eq.(\ref{403}), Eq.(\ref{401}) becomes 
\begin{equation}
 \cl_{jj}=\sum_{A=3,8}\left[
 -\frac{1}{2}j_{\mu}^A\frac{1}{\square + m_A^2}j^{A \mu}
   -\frac{1}{2}j_{\mu}^A\frac{m_A^2}{\square + m_A^2}\frac{n\cdot n}{(n\cdot\ptl)^2}
\left(g^{\mu\nu}-\frac{n^{\mu}n^{\nu}}{n\cdot n}\right)j_{\nu}^A\right].   \lb{404}
\end{equation}

     Although we used the dual electric potential $\mcb^A_{\mu}$ above, we can use 
the electric potential  $\tilde{B}^A_{\mu}$.  The relation between $\tilde{B}^A_{\mu}$ and $\mcb^A_{\mu}$ is \cite{hs19}
\begin{align}
 -\epsilon^{\mu\nu\alpha\beta}\ptl_{\alpha}\mcb^A_{\beta}&=(\ptl\wedge \bt^A)^{\mu\nu} +\Lambda_e^{A\mu\nu}, \lb{405} \\
 \Lambda_e^{A\mu\nu}&= -\frac{n^{\mu}}{n\cdot\ptl}\ptl_{\sigma}(\ptl\wedge \bt^A)^{\sigma\nu}
     +\frac{n^{\nu}}{n\cdot\ptl}\ptl_{\sigma}(\ptl\wedge \bt^A)^{\sigma\mu}.  \nonumber
\end{align}
The dual potential $\mcb^A_{\mu}$ has the electric correspondent of the Dirac string, which we call the electric string.  
The term 
$\Lambda_e^{A\mu\nu}$ represents this string.  
\footnote{ In Appendix~C, as an example, we present the massless fields $\tilde{B}^A_{\mu}$ and $\mcb^A_{\mu}$ for 
a point charge, and 
show that $\Lambda_e^{A\mu\nu}$ describes the electric string.  The relation (\ref{405}) is also used to consider 
the color electric flux in Sect.~6.}    
The field $\bt^A_{\mu}$ satisfies the equation of motion 
\begin{equation*}
     (D_{m_A}^{-1})_{\mu\nu}\bt^{A\nu} - j^A_{\mu}=0,  
\end{equation*}
and the Lagrangian that is equivalent to Eq.(\ref{401}) is \cite{hs19}
\begin{equation*}
    \cl_{\mathrm{ecl}}=\sum_{A=3,8}\left\{ -\frac{1}{4} (\ptl\wedge \bt^A)^2
    +\frac{m_A^2}{2}\bt^A_{\mu}\bt^{A\mu} - \bt^A_{\mu}j^{A\mu} 
- \frac{m_A^2}{2}\bt^{A\mu}\frac{n\cdot n}{(n\cdot \ptl)^2}\left(g_{\mu\nu}-\frac{n_{\mu}n_{\nu}}{n\cdot n}\right)j^{A\nu}\right\}.  
\end{equation*}
The last term comes from the electric string.  Substituting $\bt^{A\mu}=(D_{m_A})^{\mu\nu}j_{\nu}^A$ into 
$\cl_{\mathrm{ecl}}$, we can obtain $\cl_{jj}$ in Eq.(\ref{404}).

\subsection{Potential between static charges}

     We consider the static charges $Q^A_a$ at $\bve{a}$ and $Q^A_b$ at $\bve{b}$.  Substituting the static current 
\begin{equation}
     j^A_{\mu}(x)=g_{\mu 0}\left\{Q^A_a\delta(\bve{x}-\bve{a})+Q^A_b\delta(\bve{x}-\bve{b})\right\}  \lb{406}
\end{equation}
into $\cl_{jj}$, we get the potential 
\begin{align}
   V(\bve{r})&=\sum_{A=3,8}\left\{V_{Y}^A(r)+V_{L}^A(\bve{r})\right\},  \nonumber \\ 
 V_{Y}^A(r)&=\int \frac{d^3q}{(2\pi)^3}\left(\frac{(Q^A_a)^2+(Q^A_b)^2}{2}+Q^A_aQ^A_be^{i \bve{q}\cdot \bve{r}}\right)\frac{1}{q^2+m_A^2},  \lb{407} \\ 
     V_{L}^A(\bve{r}) &=\int \frac{d^3q}{(2\pi)^3}\left(\frac{(Q^A_a)^2+(Q^A_b)^2}{2}+Q^A_aQ^A_be^{i \bve{q}\cdot \bve{r}}\right)\frac{m_A^2}{(q^2+m_A^2)q_n^2},  \lb{408}
\end{align}
where $\bve{r}=\bve{a}-\bve{b}$, $q=|\bve{q}|$ and $q_n=\bve{q}\cdot\bve{n}$.  The first (second) term in $\cl_{jj}$ leads to $V^A_Y$ ($V^A_L$).  Historically, these potentials were obtained by using the dual Ginzburg--Landau model \cite{suz, mts, sst, sst2}.  
These potentials are calculated in Appendix~D.  Assuming that $m_A$ disappears above the scale $\Lambda_c$, 
Eq.(\ref{407}) gives \cite{hs21} 
\begin{align}
  V_Y^A(r) &=Q^A_aQ^A_b\left( \frac{1}{4\pi r}-\frac{m_A^2}{2\pi^2}\int_0^{\Lambda_c} dq \frac{\sin qr}{qr}\frac{1}{q^2+m_A^2}\right) 
\nonumber \\
 &=-\frac{Q_a^AQ_b^A}{g^2}\left(-\frac{\alpha^A(r)}{r}\right),\quad \alpha^A(r)=\frac{g^2}{4\pi}-
\frac{g^2m_A^2 r}{2\pi^2}\int_0^{\Lambda_c} dq \frac{\sin qr}{qr}\frac{1}{q^2+m_A^2}.  \lb{409}
\end{align}
The first term in $V_Y^A(r)$ is the usual Coulomb potential, which is the main term for small $r$.  

    Under the same assumption that $m_A=0$ above $\Lambda_c$, Eq.(\ref{408}) gives       
\begin{align}
V_L^A(\bve{r})=&V_{\mathrm{IR}}^A(r_t)- \frac{Q^A_aQ^A_bm_A^2}{4\pi}K_0(m_Ar_t, \Lambda_c) r_n +\cdots, \lb{410} \\ 
 V_{\mathrm{IR}}^A(r_t)=&
 \frac{m_A^2}{2\pi^2 \ve}\left\{\frac{(Q^A_a)^2+(Q^A_b)^2}{2m_A} \tan^{-1}\frac{\Lambda_c}{m_A}
+
 Q^A_aQ^A_b H(m_A,\Lambda_c, r_t)\right\},  \lb{411} 
\end{align}
where the functions $K_0(m_A r_t,\Lambda_c)$ and $H(m_A,\Lambda_c, r_t)$ are 
defined in Eq.(\ref{d11}).  
We have chosen $\bve{n}$ as $r_n=\bve{r}\cdot \bve{n} \geq 0$, and $\bve{r}=(r_n, \bve{r}_t)$.  The vector $\bve{r}_t$ 
satisfies $\bve{r}_t\bot \bve{n}$, and $r_t=|\bve{r}_t|$.  The term $V_{\mathrm{IR}}^A$ has infrared divergence 
$1/\ve$, where the infrared cut-off $\ve$ satisfies $0<\ve \ll 1$.     
To remove this divergence, since the direction $\bve{n}$ of the electric string 
is arbitrary, we choose $\bve{r}\parallel \bve{n}$ \cite{sst2, hs19, hs21}.  
In this case, as $(r_n, r_t)=(r, 0)$, Eq.(\ref{410}) becomes 
\begin{align}
   V_L^A(r)&=V_{\mathrm{IR}}^A - \frac{Q_a^AQ_b^A}{g^2}\sigma^A r +\cdots, \quad 
\sigma^A= \frac{g^2 m_A^2}{8\pi}\ln\left(\frac{\Lambda_c^2+m_A^2}{m_A^2}\right), \lb{412} \\
  V_{\mathrm{IR}}^A&=
 \frac{m_A}{4\pi^2 \ve}(Q^A_a+Q^A_b)^2 \tan^{-1}\frac{\Lambda_c}{m_A}, \lb{413}  
\end{align}
where $ K_0(0,\Lambda_c)$ and $H(m_A, \Lambda_c,0)$ are presented in Eq.(\ref{d12}).  
Eq.(\ref{413}) shows that $V_{\mathrm{IR}}^A$ vanishes if $Q_a^A+Q_b^A=0$.  
Therefore, the conditions to remove the infrared divergence are  
\begin{equation} 
 r_t=0,\quad Q^A_a+Q^A_b=0.  \lb{414}  
\end{equation}
When Eq.(\ref{414}) holds, the leading term of Eq.(\ref{412}) is the linear potential 
$-(Q_a^AQ_b^A/g^2)\sigma^A r$, which is the main term for large $r$.  

     We note the infrared divergence implies the existence of the electric string with infinite length and the mass $m_A$.  
The relation between the conditions in Eq.(\ref{414}) and the length of the electric string are depicted in Fig.~2.      

\begin{figure}
\begin{center}
\includegraphics{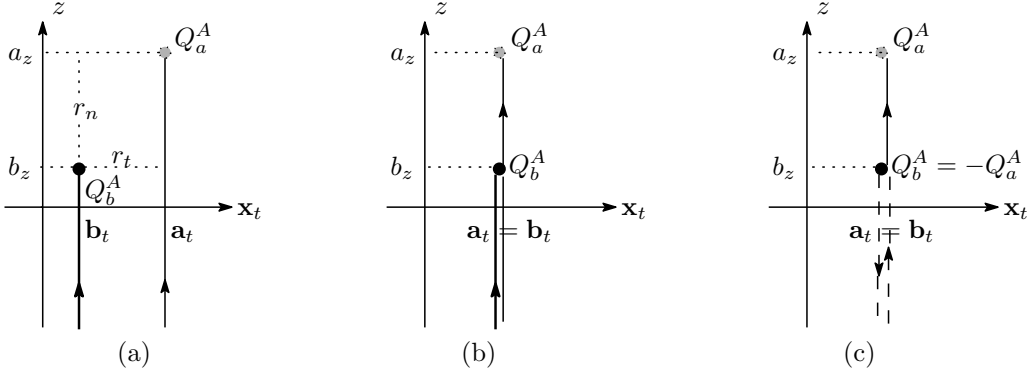}
\caption{The relation between the conditions in Eq.(\ref{414}) and the length of the electric string.  
The case (a) with $r_t\neq 0$ and the case (b) with $(r_t=0, Q_a^A+Q_b^A\neq 0)$ have the string with infinite length.  
The length of the string in (c), which satisfies Eq.(\ref{414}), is finite.}
\label{fig2}
\end{center}
\end{figure}

     In the SU(2) case, comparing the $q\bar{q}$ potential with $V_Y(r)$ and $V_L(r)$, we tried 
to determine the values of parameters, and reproduce the Coulomb plus linear type potential \cite{hs21}.  
However, in the SU(3) case, there are many parameters like $M_A\ (A=3,8)$ and $\mathcal{M}_{\alpha}\ (\alpha=1,2,3)$.  
In addition, since $m_3\neq m_8$, we are not sure whether a single cut-off $\Lambda_c$ is usable or not.  
So we don't try to determine the parameters in this paper.  Instead, we study the consequences derived from the 
Lagrangian (\ref{401}) and the potentials (\ref{409}) and (\ref{412}), below.

\section{Mesonic and baryonic potentials}

\subsection{Notation}

     Corresponding to the three types of the color charge red, blue and green, we use $C_1$, $C_2$ and $C_3$, respectively.  
     The quark field is $\Psi = \ ^t (\psi_{C_1}\ \psi_{C_2}\ \psi_{C_3})$, and the current  
$j_{\mu}^A=g\bar{\Psi}\gamma_{\mu}H^A\Psi$ $(A=3,8)$ is written as 
\begin{equation*}
     j_{\mu}^A=\sum_{i=1}^3 g w_i^A  \bar{\psi}_{C_i} \gamma_{\mu}\psi_{C_i},  
\end{equation*}
where the weight vectors are
\begin{equation}
  \vec{w}_1=\left(\frac{1}{2},\frac{1}{2\sqrt{3}}\right),\quad  \vec{w}_2=\left(\frac{-1}{2},\frac{1}{2\sqrt{3}}\right), 
\quad \vec{w}_3=\left(0,-\frac{1}{\sqrt{3}}\right).  \lb{501}
\end{equation}

     When we use the static potentials (\ref{409}) -- (\ref{413}), the static charges are given by 
\begin{equation}
  Q_{C_j}^A=gw_j^A=-\bar{Q}_{C_j}^A \quad (A=3,8,\ j=1,2,3).  
\lb{502}
\end{equation}

\subsection{Mesonic potential}

     If a static quark (an antiquark) exists at $\bve{a}$ ($\bve{b}$), a meson is expressed by 
\[ \frac{1}{\sqrt{3}} \sum_{i=1}^3 |q_{C_i}(\bve{a}) \bar{q}_{C_i}(\bve{b})\rangle   \]
We set $Q_a^A=Q_{C_i}^A$, $Q_b^A=\bar{Q}_{C_i}^A=-Q_{C_i}^A$ and
$\bve{r}=(\bve{a}-\bve{b}) \parallel \bve{n}$.  Then the two conditions in Eq.(\ref{414}) are satisfied, and $V_{\mathrm{IR}}^A$ 
vanishes.  Using the relation 
\[ \frac{1}{3}\sum_{i=1}^3 \frac{-Q_{C_i}^A \bar{Q}_{C_i}^A}{g^2}=\frac{1}{3}\sum_{i=1}^3(w_i^A)^2=\frac{1}{6}\quad  (A=3, 8),  \]
Eqs.(\ref{409}) and (\ref{412}) give the mesonic potential 
\begin{equation}
  V_{q\bar{q}}(r)=\frac{1}{6}\sum_{A=3,8} \left\{-\frac{\alpha^A(r)}{r}+\sigma^A r +\cdots \right\}=
-\frac{\alpha_{q\bar{q}}(r)}{r}+\sigma_{q\bar{q}} r +\cdots,  
\lb{503}
\end{equation}
where $\alpha_{q\bar{q}}(r)=\sum_{A=3,8} \alpha^A(r)/6$ and $\sigma_{q\bar{q}}=\sum_{A=3,8}\sigma^A/6$.  

\begin{figure}
\begin{center}
\includegraphics{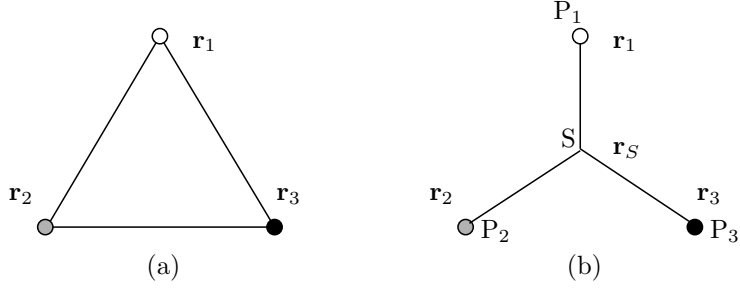}
\caption{The color flux between the charges. Fig.~(a) is the $\Delta$--type baryon and Fig.~(b) is the $Y$--type baryon.  }
\label{fig3}
\end{center}
\end{figure}

\subsection{$\Delta$-type $3q$ potential}

     Let us study the potential for the configuration in Fig.~3(a), which is called the $\Delta$-ansatz \cite{cw}.  
To apply Eqs.(\ref{409}) and (\ref{412}), 
we replace $r$ with $r_{kl}=|\bve{r}_{kl}|=|\bve{r}_k-\bve{r}_l|\ (k\neq l)$, and $\bve{n}$ with 
$\bve{n}_{kl}$ which satisfies  $\bve{n}_{kl} \parallel \bve{r}_{kl}$.  
When static quarks are placed at $\bve{r}_k$ $(k=1,2,3)$, a baryonic state is 
\[ \frac{1}{\sqrt{6}}\sum_{ijk} \varepsilon_{ijk}|q_{C_i}(\bve{r}_1)q_{C_j}(\bve{r}_2)q_{C_k}(\bve{r}_3)\rangle.  \]
If we set $Q_a^A=Q_{C_i}^A$ and $Q_b^A=Q_{C_j}^A$ with $i\neq j$, and use 
the relation 
\[  \frac{1}{6}\sum_{i\neq j} \frac{-Q_{C_i}^A Q_{C_j}^A}{g^2}=-\frac{1}{6}\sum_{i\neq j}w_i^Aw_j^A
=\frac{1}{12} \quad  (A=3, 8),  \]
Eqs.(\ref{409}) and (\ref{412}) give 
\begin{equation}
  V_{3q}^{\Delta}(\bve{r}_1,\bve{r}_2,\bve{r}_3)=\sum_{A=3,8}V_{\mathrm{IR}}^A + 
\frac{1}{12}\sum_{k>l}\sum_{A=3,8} \left\{-\frac{\alpha^A(r_{kl})}{r_{kl}}+\sigma^A r_{kl} +\cdots,  \right\}.  \lb{504}
\end{equation}

     We make two comments.  First, 
from Eqs.(\ref{503}) and (\ref{504}), we obtain the relation \cite{afj}
\begin{equation}
    V_{3q}^{\Delta}(\bve{r}_1,\bve{r}_2,\bve{r}_3)-\sum_{A=3,8}V_{\mathrm{IR}}^A =  \frac{1}{2}\sum_{k>l}V_{q\bar{q}}(r_{kl}).  
\lb{505} 
\end{equation}
Second, by the choice $\bve{n}_{kl} \parallel \bve{r}_{kl}$, the first condition $(r_{kl})_t =0$ is satisfied.  
However, 
except for $Q^3_{C_1}+Q^3_{C_2}=0$, the second condition $Q^A_{C_i}+Q^A_{C_j}= 0\ (i\neq j)$ does not hold.  
So, using 
\[   \frac{1}{6}\sum_{i\neq j} \frac{(Q_{C_i}^A + Q_{C_j}^A)^2}{g^2}=\frac{1}{6}\sum_{i\neq j}(w_i^A + w_j^A)^2
=\frac{1}{6} \quad  (A=3, 8),  \] 
we find the infrared divergent term  
\begin{equation}
     \sum_{A=3,8}V_{\mathrm{IR}}^A= \sum_{A=3,8} \frac{m_A}{24\pi^2 \ve} \tan^{-1}\frac{\Lambda_c}{m_A}   \lb{506}
\end{equation}
remains.  In the $\Delta$-ansatz, there are electric strings with infinite length.  When $m_A\neq 0$, they give rise to the 
infrared divergence.

\subsection{$Y$-type $3q$ potential}

     For large $r_{kl}$, the potential $V_L^A(r_{kl})$ in 
$V_{3q}^{\Delta}(\bve{r}_1,\bve{r}_2,\bve{r}_3)$ has the infrared divergence.  
On the other hand, based on the strong coupling argument, the $Y$-shaped baryon depicted in Fig.~3(b)  was proposed \cite{ckp}.  
The point S at $\bve{r}_S$, where the sum of the length 
$L_Y=\sum_{k=1}^3 r_{kS}=\sum_{k=1}^3|\bve{r}_k-\bve{r}_S|$ becomes minimum,  
is the Steiner point.  The color electric flux lines emanating from the three quarks meet 
and disappear there.  Since the state at this point is color singlet, corresponding to the state 
$|q_{C_1}(\bve{r}_1)q_{C_2}(\bve{r}_2)q_{C_3}(\bve{r}_3)\rangle$, the state at $\bve{r}_S$ is 
$|\bar{q}_{C_1}(\bve{r}_S)\bar{q}_{C_2}(\bve{r}_S)\bar{q}_{C_3}(\bve{r}_S)\rangle$.  
So, when $r_{kS}$ is large, the potential is the sum of the three $q\bar{q}$ potentials for large $r$.     
Thus we obtain  
\begin{equation}
   V_{3qL}^{Y}(\bve{r}_1,\bve{r}_2,\bve{r}_3)= \sum_{k=1}^3V_{q\bar{q}L}(r_{kS})=
 \frac{1}{6}\sum_{k=1}^3 \sum_{A=3,8} (\sigma^Ar_{kS} + \cdots) = \sigma_Y L_Y + \cdots,   \lb{507}
\end{equation}
where $\sigma_Y= \sum_{A=3,8} \sigma^A/6$.  

     We note, when $r_{kl}$ is large, Eq.(\ref{504}) gives 
\begin{equation}
  V_{3qL}^{\Delta}(\bve{r}_1, \bve{r}_2,\bve{r}_3)-\sum_{A=3,8}V_{\mathrm{IR}}^A 
=\sigma_{\Delta}L_{\Delta} +\cdots,  \lb{508} 
\end{equation}  
where $L_{\Delta}= \sum_{k>l} r_{kl}$ and $\sigma_{\Delta}=\sum_{A=3,8} \sigma^A/12$.  
From Eqs.(\ref{503}),  (\ref{507}) and (\ref{508}), the relations  $\sigma_{\Delta}= \sigma_{q\bar{q}}/2$,   
$\sigma_{Y}=\sigma_{q\bar{q}}$ and 
\begin{equation}
     V_{3qL}^{\Delta}(\bve{r}_1, \bve{r}_2,\bve{r}_3)-\sum_{A=3,8}V_{\mathrm{IR}}^A =  V_{3qL}^{Y}(\bve{r}_1, \bve{r}_2,\bve{r}_3) 
-\sigma_{q\bar{q}}\left(L_Y-\frac{1}{2}L_{\Delta}\right)   \lb{509}
\end{equation}
are obtained at this level \cite{afj}.  
As $L_Y>L_{\Delta}/2$, the inequality $V_{3qL}^Y>V_{3qL}^{\Delta}-\sum_{A=3,8}V_{\mathrm{IR}}^A$ holds.  
However, different from $V_{3qL}^{\Delta}$, $V_{3qL}^Y$ is free from the infrared divergence.

\begin{figure}
\begin{center}
\includegraphics{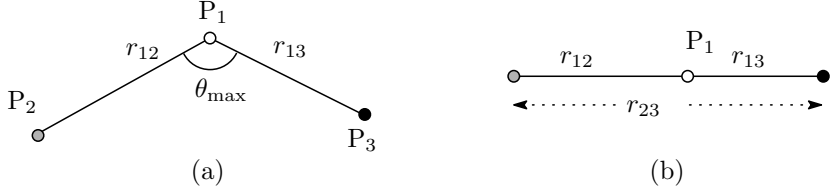}
\caption{The cases that the maximum inner angle $\theta_{\mathrm{max}}$ satisfies 
$120^{\circ}\leq \theta_{\mathrm{max}}\leq 180^{\circ}$.   }
\label{fig4}
\end{center}
\end{figure}

\subsection{Comparison with the lattice results}

     In the present model, the classical Abelian potentials $\mathcal{B}_{\mu}^A$ $(A=3,8)$ lead to the linear potential.  
The $Y$-type potential is preferable to the $\Delta$-type potential, because the former has no infrared divergence.  
The string tension of the $Y$-type potential satisfies $\sigma_Y=\sigma_{q\bar{q}}$.  
     
     In the lattice simulation, the $3q$ baryon has been studied, and the $Y$-type potential is obtained \cite{bi,sasu,kk}.  
In Ref.~\cite{sasu}, using the maximal Abelian gauge, it is shown that the three quark string tension $\sigma_{3q}$ 
satisfies $\sigma_{3q}\simeq \sigma_{q\bar{q}}$.  In addition, the string tensions 
$\sigma_{3q}^{\mathrm{Abel}}$ and $\sigma_{q\bar{q}}^{\mathrm{Abel}}$, which are obtained from the Abelian part, 
satisfy $\sigma_{3q}\simeq \sigma_{3q}^{\mathrm{Abel}}$ and $\sigma_{q\bar{q}}\simeq \sigma_{q\bar{q}}^{\mathrm{Abel}}$ 
within a few percent deviation.  These results show that the potential is $Y$-type, and the Abelian dominance is realized.  

     In Ref.~\cite{kk}, using the Polyakov loop correlation function, 
the cases with 
$60^{\circ}\leq \theta_{\mathrm{max}}< 120^{\circ}$ and $120^{\circ}\leq \theta_{\mathrm{max}}\leq 180^{\circ}$ 
are simulated, where $\theta_{\mathrm{max}}$ represents the maximum inner angle of a triangle.   
In the latter case, the Steiner point $\mathrm{S}$ is the point $\mathrm{P}_1$ in Fig.~4.  
As $r_{1S}=0$ in this case, the length $L_Y$ is reduced to $L_Y=r_{12}+r_{13}=L_{\Delta}-r_{23}$.  
When $120^{\circ}\leq \theta_{\mathrm{max}}< 180^{\circ}$, they found that the long-range potential satisfies 
$V_{3qL}\simeq \sigma_{q\bar{q}}L_Y$, and $\sigma_{3q}\simeq \sigma_{q\bar{q}}$ holds.   
On the other hand, when $\theta_{\mathrm{max}}= 180^{\circ}$, they obtained the $\Delta$-type relation 
$\displaystyle V_{3q}=\frac{1}{2}\sum_{k> l}V_{q\bar{q}}(r_{kl})$.  

     In our approach, when $120^{\circ}\leq \theta_{\mathrm{max}}\leq 180^{\circ}$, the $Y$-type potential 
is calculable by setting $r_{1S}=0, r_{2S}=r_{12}$ and $r_{3S}=r_{13}$.  The result is 
\begin{equation}
     V_{3qL}^Y=\sigma_Y L_Y = \sigma_{q\bar{q}}L_Y ,\quad L_Y=r_{12}+r_{13}.  \label{510}
\end{equation}
When $\theta_{\mathrm{max}}= 180^{\circ}$, Fig.~4(b) shows $r_{23}=r_{12}+r_{13}$ and $L_Y=L_{\Delta}/2$.  
As $\sigma_{\Delta}=\frac{1}{2}\sigma_{Y}$, we find Eq.(\ref{510}) becomes 
\begin{equation}
     \sigma_{\Delta}L_{\Delta}= \frac{1}{2}\sigma_{q\bar{q}}L_{\Delta} ,\quad L_{\Delta}==r_{12}+r_{13}+r_{23}=2L_Y.  \label{511}
\end{equation}
Namely, if $\theta_{\mathrm{max}}= 180^{\circ}$, the $Y$-type 
relation Eq.(\ref{510}) coincides with the $\Delta$-type relation (\ref{511}), which is expected from Eq.(\ref{509}).  
Therefore we can say that the long-range potential is $Y$-type for $\theta_{\mathrm{max}}\leq 180^{\circ}$.

\section{Color electric flux}

\subsection{Extended Maxwell's equations}

     In Sect.~4, we introduced the electric potential $\bt^A_{\mu}$ and its dual potential $\mcb^A_{\mu}$ that are related 
by the Eq.(\ref{405}).  We also used the Zwanziger's dual field strength $^dF^{A\mu\nu}$ \cite{zwa} in the presence of 
the current $j^A_{\mu}$.  In this subsection, we study the Maxwell's equations.  

     Using  $\mcb^A_{\mu}$ and $\bt^A_{\mu}$, the dual field strength is expressed by  
\begin{align*}
^dF^{A\mu\nu}=&(\ptl\wedge \mcb^A)^{\mu\nu}+\frac{1}{n\cdot\ptl}\epsilon^{\mu\nu\alpha\beta}n_{\alpha}j^A_{\beta}\\
=&\epsilon^{\mu\nu\alpha\beta}\ptl_{\alpha}\bt^A_{\beta} -\frac{1}{n\cdot\ptl}\epsilon^{\mu\nu\alpha\beta}n_{\alpha}\left\{\ptl\cdot(\ptl\wedge\bt^A)-j^A\right\}_{\beta},   
\end{align*} 
and the field strength is 
\begin{align*}
 F^{A\mu\nu}=&-\epsilon^{\mu\nu\alpha\beta}\ptl_{\alpha}\mcb^A_{\beta}+\frac{1}{n\cdot \ptl}(n\wedge j^A)^{\mu\nu}\\
=&(\ptl\wedge \bt^A)^{\mu\nu} -\frac{1}{n\cdot\ptl}\left[n\wedge \left\{\ptl\cdot(\ptl\wedge \bt^A)-j^A\right\}\right]^{\mu\nu}.  
\end{align*}
The electric field $E^{Ai}(j)=F^{Ai0}$ and the magnetic field $H^{Ai}(j)=\ ^dF^{Ai0}$
are  
\begin{align}
E^{Ai}(j)=&-\epsilon^{ijk}\ptl_j\mcb^{Ak}+\frac{n^i}{n\cdot \ptl}j^{A0}  \nonumber \\
=&-\ptl_i \bt^{A0}-\ptl_0\bt^{Ai}-\frac{n^i}{n\cdot\ptl}\left\{\ptl\cdot(\ptl\wedge\bt^A)-j^A\right\}^0,   \lb{601}  \\
H^{Ai}(j)=&-\ptl_i \mcb^{A0} -\ptl_0\mcb^{Ai} - \epsilon^{ijk}\frac{n^j}{n\cdot\ptl}j^{Ak}  \lb{602} \\
=& \epsilon^{ijk}\ptl_j\bt^{Ak}+\epsilon^{ijk}\frac{n^j}{n\cdot\ptl}\left\{\ptl\cdot(\ptl\wedge \bt^A)-j^A\right\}^k, 
\nonumber
\end{align}
where $n^0=0$ has been used.  
From these expressions, it is easy to show the two Maxwell's equations  
\begin{equation}
  \nabla \cdot \bve{E}^{A}(j)=j^{A0}, \quad \nabla \times \bve{H}^{A}(j)-\ptl_0 \bve{E}^{A}(j)=\bve{j}^{A}. \lb{603}
\end{equation}

     Next we consider the remaining two Maxwell's equations.  
Using $\mcb^{A\mu}$, Eq.(\ref{602})  gives 
\begin{align*}
  \ptl_i H^{Ai}(j)=& \ptl_i\left( -\ptl_i \mcb^{A0} -\ptl_0\mcb^{Ai} - \epsilon^{ijk}\frac{n^j}{n\cdot\ptl}j^{Ak}\right)  \\
  =& \ptl\cdot (\ptl\wedge \mcb^A)^0-\mathcal{J}^{A0}, \quad 
\mathcal{J}^{A\mu}=\frac{1}{n\cdot \ptl}\epsilon^{\mu\nu\alpha\beta}\ptl_{\nu}n_{\alpha}j^A_{\beta}.  
\end{align*}
Since the classical fields $\mcb^{A\mu}$ satisfies the equation of motion 
\begin{equation}
   \ptl\cdot (\ptl\wedge \mcb^A)^{\mu}  + m_A^2\mcb^{A\mu}=\mathcal{J}^{A\mu},  \lb{604}  
\end{equation}
the above equation becomes 
\begin{equation}
   \nabla \cdot \bve{H}^{A}(j)= -m_A^2\mcb^{A0}.    \lb{605}
\end{equation}
In the same way, we obtain 
\begin{equation}
  -\nabla\times \bve{E}^{A}(j)-\ptl_0 \bve{H}^{A}(j) = -m_A^2\bve{\mcb}^{A}.  \lb{606}
\end{equation}
Namely, because of the term $-m_A^2\mcb^{A\mu}=-m_A^2(\mcb^{A0}, \bve{\mcb}^{A})$, 
the remaining two Maxwell's equations are modified.  

     If we consider a model with the magnetic current 
$j_{\mathrm{mag}}^{A\mu}=(\rho_{\mathrm{mag}}^{A}, \bve{j}_{\mathrm{mag}}^{A})$, 
$\rho_{\mathrm{mag}}^{A}$ and $\bve{j}_{\mathrm{mag}}^{A}$ will appear in the right hand side of Eqs.(\ref{605}) and (\ref{606}), 
respectively.   In the dual superconductor model, there is the monopole field, and 
the static equation $ -\nabla\times \bve{E}^{A}(j) = \bve{j}_{\mathrm{mag}}^A$ is often discussed \cite{rip, kst}.  
In the present model, there is no monopole field and no magnetic current originally.  
However, like the London equation in superconductivity, 
the relation $\bve{j}_{\mathrm{mag}}^A=-m_A^2\bve{\mcb}^{A}$ appears.

\subsection{Color flux tube}

     It is expected that the color flux tube connects color charges.  In Ref.~\cite{kst}, using the dual superconductor model, 
the color flux is studied.  From this flux and the equation 
$ -\nabla\times \bve{E}^{A}(j) = \bve{j}_{\mathrm{mag}}$, the magnetic current is also investigated.  In this subsection, 
we consider the color flux tube.   

     Let us consider the electric flux between the charges $Q^A_{C_i}$ at $\bve{a}=(0,0,a)$ and 
$\bar{Q}^A_{C_i}$ at $\bve{b}=(0,0,b)$.  
We set  $\bve{n}=(0,0,1)$, and assume that the mass $m^A$ is approximately constant for $\rho>1/\Lambda_c$, where 
$(\rho, \theta, z)$ are the cylindrical coordinates.  
To study the static flux tube solution, we set $\mcb^{A0}=0$ and 
\begin{equation}
   \bve{\mcb}^{A}(\rho, \theta, z) \approx B(\rho)f(z)\bve{e}_{\theta},  \lb{607}
\end{equation}     
where the unit vectors are 
\[
 \bve{e}_{\rho}=(\cos \theta, \sin \theta, 0), \quad \bve{e_{\theta}}=(-\sin \theta, \cos \theta, 0),\quad \bve{e}_z=(0,0,1).  
\]
Substituting Eq.(\ref{607}) into Eq.(\ref{604}), we obtain 
\begin{equation}
    \left(\frac{\ptl^2}{\ptl \rho^2}+\frac{1}{\rho}\frac{\ptl}{\ptl \rho}-\frac{1}{\rho^2} - m_A^2\right)B(\rho)f(z)
+B(\rho)f''(z) = \frac{1}{\ptl_z}\frac{\ptl}{\ptl \rho} j^A_0.   \lb{608} 
\end{equation}
Since $j^A_0=0$ holds for $\rho>0$, if we assume $f''(z)\approx 0$ in the interval $b<z<a$, Eq.(\ref{608}) reduces to 
the equation 
\[   \left(\frac{\ptl^2}{\ptl \rho^2}+\frac{1}{\rho}\frac{\ptl}{\ptl \rho}-\frac{1}{\rho^2} - m_A^2\right)B(\rho)\approx 0 \]
in the region $(\rho>1/\Lambda_c, b<z<a)$.  
The solution of this equation with $\lim_{\rho \to \infty} B(\rho) =0$ is \cite{no}  
\[   B(\rho) =\lambda K_1(m_A\rho),  \]
where $\lambda$ is a constant, and $K_n(X)$ is the modified Bessel function.  
So, we obtain 
\begin{equation}
      \bve{\mcb}^{A}\approx \lambda K_1(m_A\rho)f(z)\bve{e}_{\theta}.  \lb{609}
\end{equation}
Using Eq.(\ref{609}) and the equality $XK_n'(X)+nK_n(X)=XK_{n-1}(X)$, the color electric field becomes 
\begin{equation}
   \bve{E}^{A}(j)=-\nabla\times \bve{\mcb}^A \approx m_A \lambda K_0(m^A\rho)f(z)\bve{e}_z
 + \lambda K_1(m_A\rho)f'(z)\bve{e}_{\rho}.  \lb{610}
\end{equation}
In the same way, if we apply the relations  $XK_n'(X)-nK_n(X)=-XK_{n+1}(X)$ and $K_0'(X)=-K_1(X)$, we get   
\[ m^A \lambda \frac{\ptl}{\ptl \rho}K_0(m^A\rho) f(z) (\bve{e}_\rho\times \bve{e}_z) 
= -m_A^2 \lambda K_1(m_A\rho) f(z) \bve{e}_{\theta} = -m_A^2\bve{\mcb}^A.  
\]
From this equation and Eq.(\ref{610}), we obtain 
\begin{equation}
   -\nabla\times \bve{E}^{A}(j)= -m_A^2\bve{\mcb}^A +\lambda K_1(m^A\rho)f''(z)\bve{e}_{\theta}.  \lb{611}
\end{equation}
In the interval $b<z<a$, $f''(z)\approx 0$ is assumed, and Eq.(\ref{611}) becomes Eq.(\ref{606}) with $\ptl_0\bve{H}^A=0$.

\subsection{Flux tube represented by $\bt^{A\mu}$}

    Next, we restudy the flux tube by using the electric potential $\bt^{A\mu}$.  In the static case, Eq.(\ref{601}) becomes  
\begin{equation}
  \bve{E}(j)^A=-\nabla \bt^{A0}-\frac{\bve{n}}{n\cdot \ptl}\Box \bt^{A0}.  \lb{612}
\end{equation}
From the equation of motion 
\[  \ptl\cdot (\ptl\wedge \bt^A)^{\mu}  + m_A^2\bt^{A\mu}=j^{A\mu},  \]
$\bt^{A0}$ satisfies 
\[  \nabla^2 \bt^{A0}-m_A^2\bt^{A0}=-j^{A0}.  \]
If we can write $\bt^{A0}\approx D(\rho)h(z)$ approximately, this equation becomes  
\begin{equation} 
     \left(\frac{\ptl^2}{\ptl \rho^2}+\frac{1}{\rho}\frac{\ptl}{\ptl \rho}-m_A^2\right)D(\rho)h(z)+D(\rho)h'(z)=-j^{A0}.  \lb{613}
\end{equation}
As in the previous subsection, we set $j^{A0}=0$ for $\rho>0$, and assume $h'(z)\approx 0$ in the interval 
$b<z<a$.  
Then Eq.(\ref{613}) becomes 
\[   \left(\frac{\ptl^2}{\ptl \rho^2}+\frac{1}{\rho}\frac{\ptl}{\ptl \rho}-m_A^2\right)D(\rho)\approx 0.  \]
Using the constant $\kappa$, the solution of this equation is 
\[  D(\rho)=\kappa K_0(m_A\rho).  \]
Since we choose $\bve{n}=(0,0,1)$,  $\bt^{A0}\approx \kappa K_0(m_A\rho)h(z)$ 
gives 
\begin{align}
 -\nabla \bt^{A0}&=m_A\kappa K_1(m_A\rho)h(z) \bve{e}_{\rho}- \kappa K_0(m_A\rho)h'(z)\bve{e}_z, \nonumber \\
 -\frac{\bve{n}}{n\cdot \ptl}\Box \bt^{A0} &= \kappa m_A^2K_0(m_A\rho)\frac{1}{\ptl_z}h(z) \bve{e}_z + 
\kappa K_0(m_A\rho)h'(z)\bve{e}_z,  \lb{614}
\end{align}
and Eq.(\ref{612}) becomes  
\begin{equation}
  \bve{E}(j)^A=\kappa m_A^2K_0(m_A\rho)\frac{1}{\ptl_z}h(z) \bve{e}_z + m_A\kappa K_1(m_A\rho)h(z) \bve{e}_{\rho}.  \lb{615}
\end{equation}
Comparing Eqs.(\ref{610}) and (\ref{615}), we can identify 
\footnote{The minus sign comes from the choice that the electric string is in the negative $z$-direction.  See Eq.(\ref{c04}).} 
\[  \kappa m_A=-\lambda, \quad -\frac{1}{\ptl_z}h(z)=f(z),\quad -h(z)=f'(z).  \]
So, $f(z)$ and $h(z)$ can be approximated by  
\[ f(z)\approx \theta(a-z)-\theta(b-z),\quad h(z)\approx \delta(z-a)-\delta(z-b),  \]
where $\theta(z)$ is the unit step function.

     Thus, the electric potential 
\begin{equation}
   \bt^{A0}\approx -\frac{\lambda}{m_A}K_0(m_A\rho)\left\{\delta(z-a)-\delta(z-b)\right\}  \lb{616}
\end{equation}
produces the color electric flux 
\begin{equation}
   \bve{E}^A(j)\approx m_A \lambda K_0(m_A\rho)\left\{\theta(a-z)-\theta(b-z)\right\}\bve{e}_z  \lb{617}
\end{equation}
in the region $(\rho>1/\Lambda_c, b<z<a)$.  The string part (\ref{614}) is responsible for this flux tube.  
The corresponding dual potential is 
\begin{equation}
   \bve{\mcb}^A\approx \lambda K_1(m_A\rho)\left\{\theta(a-z)-\theta(b-z)\right\}\bve{e}_{\theta},  \lb{618}
\end{equation}
which also gives the flux (\ref{617}).  This flux satisfies the extended Maxwell's equation 
\begin{equation}   
  -\nabla\times \bve{E}^A(j)\approx -m_A^2\bve{\mcb}^A, \lb{619} 
\end{equation}
where the magnetic current is $\bve{j}_{\mathrm{mag}}^A=-m_A^2\bve{\mcb}^A$.

    We make a comment.  The lattice simulation shows that the $3q$ baryon 
is $Y$-shaped \cite{bi, sasu, kk} and 
the solenoidal magnetic current exists \cite{bi}.  In the present approach, the $Y$-type baryonic potential 
is free from infrared divergence, and it consists of three $q\bar{q}$ potentials.  So, although the flux tube of 
$q\bar{q}$ is considered here, we can apply it to the  $Y$-type $3q$ baryon.  
The flux tube of $q\bar{q}$ can exist between $\bve{r}_S$ and $\bve{r}_k\ (k=1,2,3)$.  
The current $\bve{j}_{\mathrm{mag}}^A=-m_A^2\bve{\mcb}^A$ with (\ref{618}), which has the solenoidal form, 
also appears.  .

\section{Summary and comment}

     In the dual superconductor picture of the quark confinement, the monopole condensation produces the gluon mass.  
To realize this scenario, the dual Ginzburg--Landau model introduces the monopole field, and its condensation, 
the gluon mass and the static potential have been studied.  

     In Ref.~\cite{hs19}, we considered another possibility to make Abelian component of the gluon massive 
in the SU(2) gauge theory.  The static potential was also studied \cite{hs21}.  
In this paper, we extended this approach to the SU(3) gauge theory.   
In the nonlinear gauge of the Curci--Ferrari type, quartic ghost interaction generates the ghost condensate  
$v^A=g\langle \vp^A \rangle$ below the scale $\Lambda_{\mathrm{QCD}}$.  The ghost loop with $v^A$ gives rise to the 
tachyonic mass to the quantum part of the gluon.  This tachyonic mass is removable by the gluon condensate 
$\langle A_{\mu}^aA^{a\mu}\rangle$.  Since the classical part $b^A_{\mu}$ of the gluon has no tachyonic mass, 
the condensate $\langle A_{\mu}^aA^{a\mu}\rangle$ gives the mass $m_A$ to this part.  To study the color confinement, 
the dual color electric potential $\mcb^A_{\mu}$, which is equivalent to the color electric potential $\bt^A_{\mu}$ with 
the string part $\Lambda_e^{A\mu\nu}$, was chosen as $b^A_{\mu}$.  Thus, the classical Lagrangian we use is 
$\mathcal{L}_{\mathrm{cl}}$ in Eq.(\ref{401}).  

    This Lagrangian becomes $\mathcal{L}_{jj}$ in Eq.(\ref{404}), and it gives 
the static potential between the charges $Q_a^A$ and $Q_b^A$  
with distance $r$.  
When $r$ is small, the leading term is $V_Y^A(r)$ in Eq.(\ref{409}).  For large $r$, $V_L^A(\bve{r})$ in Eq.(\ref{410}) 
is the main term.  However, $V_L^A(\bve{r})$ contains the infrared divergence $V_{\mathrm{IR}}^A(r_t)$, 
which comes from the mass $m_A$ and the electric string with infinite length.  If the conditions $r_t=0$ and 
$Q_a^A+Q_b^A=0$ in Eq.(\ref{414}) are fulfilled, $V_{\mathrm{IR}}^A(r_t)$ vanishes.  
In this case, $V_L^A(r)$ becomes the linear potential in Eq.(\ref{412}).  

     We stress the derivation of the Lagrangian $\mathcal{L}_{\mathrm{cl}}$ is based on the one-loop calculation.  
In addition, the constant sources $K_A$ and $\mathcal{K}_{\alpha}$ are assumed.  The mass $m_A$ in Eq.(\ref{401}) was 
also assumed to be constant below the cut-off $\Lambda_c$ and vanish above $\Lambda_c$.  
These quantities must be determined.  However, different from the SU(2) case, there are many parameters in SU(3).  
We skipped the determination in this paper, and studied the consequences of the Lagrangian $\mathcal{L}_{\mathrm{cl}}$ 
and the potential $V_L^A(\bve{r})$.

    In the $q\bar{q}$ case, the two conditions in Eq.(\ref{414}) 
are satisfied, and the static potential $V_{q\bar{q}}(r)$ in Eq.(\ref{503}) is obtained.  In the $3q$ case, 
if the $\Delta$-ansatz holds, the potential 
$V_{3q}^{\Delta}$ is given by Eq.(\ref{504}).  However, since the second condition of Eq.(\ref{414}) is not fulfilled, 
the infrared divergence (\ref{506}) remains.  Contrary to the $\Delta$-ansatz, the $Y$-ansatz satisfies the two conditions.   
The potential $V_{3qL}^{Y}$ in Eq.(\ref{507}), which is free from the infrared divergence, is expected for 
large $r$.  

    Using the color electric potential $\bt^A_{\mu}$ and its dual potential $\mcb^A_{\mu}$, the color electric field 
$\bve{E}^A$ and the magnetic field $\bve{H}^A$ were investigated.  Although they satisfy the two Maxwell's equations 
(\ref{603}), because of the mass $m_A$, the remaining two equations are modified as Eqs.(\ref{605}) and (\ref{606}).  
In the static case, Eq.(\ref{606}) becomes $ -\nabla\times \bve{E}^{A}(j) = -m_A^2\bve{\mcb}^{A}$.  
In the dual Ginzburg--Landau model, which contains the monopole field, the equation 
$ -\nabla\times \bve{E}^{A}(j) = \bve{j}_{\mathrm{mag}}$ has been discussed.  
In our model, although there is no monopole field, the current $-m_A^2\bve{\mcb}^{A}$ plays the 
role of the magnetic current $\bve{j}_{\mathrm{mag}}$.   

    It is expected that the color flux tube exists between color charges.  The dual electric potential $\bve{\mcb}^A$ in Eq.(\ref{618}) 
produces the electric flux $\bve{E}^{A}(j)$ in Eq.(\ref{617}), and they satisfy Eq.(\ref{619}).  
Namely, without the monopole field, the flux tube 
$\bve{\mcb}^A$ leads to the magnetic current $\bve{j}_{\mathrm{mag}}=-m_A^2\bve{\mcb}^{A}$.   
The corresponding electric potential $\bt^{A0}$ is 
presented in Eq.(\ref{616}).  The string part (\ref{614}) is the origin of the flux tube (\ref{617}).

     Comparing the SU(3) case with the SU(2) case in Ref.~\cite{hs19}, there are some differences.  
For example, as we stated in Sect. 3, although the condensate of the diagonal component $\langle A^{3\mu}A^3_{\mu}\rangle$ vanishes in SU(2), 
the condensates $\langle A^{A\mu}A^A_{\mu}\rangle$ ($A=3,8$) exist in SU(3).  
Eq.(\ref{311}) shows there are two different mass scales 
$\sqrt{5}m/2$ and $\sqrt{3}m/2$, and the classical electric potentials $\tilde{B}_{\mu}^3$ and $\tilde{B}_{\mu}^8$ 
have different masses, whereas the tachyonic mass term in SU(2) has one scale $m$.  

     Since we have not determined the parameters yet, it is difficult to study differences between SU(2) and SU(3) 
concretely.  In Ref.~\cite{ks2}, the differences are discussed.  One of the issues is the type of the 
dual superconductivity.  Investigating the electric flux, it is concluded that the SU(3) theory is type-I, whereas the SU(2) theory 
is weak type-I or the border between type-I and type-II.  
In Appendix E, assuming the phenomenological Lagrangian for the order parameters $\mathcal{G}^{\alpha}$ and $G^A$, 
we consider the type of dual superconductivity in the present model.  
Because of the condensate $\langle A^{8\mu}A^8_{\mu}\rangle$ and the two mass scales $\sqrt{5}m/2$ and $\sqrt{3}m/2$, 
the value of the Ginzburg-Landau parameter for SU(3) may become smaller than that for SU(2).

\appendix

\section{$\Lambda_{\mathrm{QCD}}$ and $\alpha_2$}

     In the momentum region $\mu \geq \Lambda_{\mathrm{QCD}}$, as the effective potential $V(\vp)$ in Eq.(\ref{205}) 
gives $v=0$, we consider the Wilsonian effective action 
\begin{align}
\Gamma_{[\mu,\Lambda]} &= \int d^4x \left\{\sum_{\alpha=1}^3 \left(\frac{(g\vec{\epsilon}_{\alpha}\cdot \vec{\varphi})^2}{3\alpha_2 g^2} - 
\int_{\mu}^{\Lambda} \frac{d^4k}{(2\pi)^4} \ln \left[(-k^2)^2 +(g\vec{\epsilon}_{\alpha}\cdot \vec{\varphi})^2 \right]\right)\right\}
\nonumber \\
 &= \int d^4x \left\{\sum_{\alpha=1}^3 \left(\frac{(g\vec{\epsilon}_{\alpha}\cdot \vec{\varphi})^2}{3\alpha_2 g^2} - 
\int_{\mu}^{\Lambda} \frac{d^4k}{(2\pi)^4} \frac{(g\vec{\epsilon}_{\alpha}\cdot \vec{\varphi})^2}{(-k^2)^2} + \cdots\right)\right\}
\nonumber \\
 &=\int d^4x \left\{\sum_{\alpha=1}^3 \left(\frac{1}{3\alpha_2g^2}-\frac{1}{8\pi^2}\ln \frac{\Lambda}{\mu}\right)(g\vec{\epsilon}_{\alpha}\cdot \vec{\varphi})^2 +
 \cdots \right\}.  \lb{a01}
\end{align}
If $\bar{g}$ and $\bar{\alpha}_2$ represent the quantities at the scale $\mu$, Eq.(\ref{a01}) implies 
\begin{equation}
  \frac{1}{\bar{\alpha}_2\bar{g}^2} =\frac{1}{\alpha_2g^2}-\frac{3}{8\pi^2}\ln \frac{\Lambda}{\mu}= -\frac{3}{8\pi^2}\ln \frac{\mu_0}{\mu},\quad 
\mu_0=\Lambda \exp\left(-\frac{8\pi^2}{3\alpha_2 g^2}\right).   \lb{a02}
\end{equation}
From Eq.(\ref{a02}), we obtain .
\begin{equation}
 \mu \frac{\ptl}{\ptl \mu}\bar{\alpha}_2\bar{g}^2=-\frac{3}{8\pi^2}(\bar{\alpha}_2\bar{g}^2)^2.  \lb{a03}
\end{equation}
Since $\bar{g}$ satisfies 
\begin{equation}
   \mu \frac{\ptl}{\ptl \mu}\bar{g}=-\frac{\beta_0}{(4pi)^2} \bar{g}^3, \quad \beta_0=\frac{11}{3}N \lb{a04}
\end{equation}
at the one-loop level, Eqs.(\ref{a03}) and (\ref{a04}) leads to the equation 
\begin{equation*}
  \mu \frac{\ptl}{\ptl \mu}\bar{\alpha}_2=\frac{\bar{\alpha}_2\bar{g}^2}{8\pi^2}(\beta_0-3\bar{\alpha}_2).  
\end{equation*}
Namely $\alpha_2=\beta_0/3$ is the ultraviolet fixed point \cite{hs07, ks}, i.e., 
\[ \lim_{\mu \to \Lambda}\bar{\alpha}_2=\alpha_2=\frac{\beta_0}{3}.  \]
Substituting this $\alpha_2$, we find $\mu_0=\Lambda \exp\left(-8\pi^2/\beta_0g^2\right)=\Lambda_{\mathrm{QCD}}$ \cite{hs07, ks}.

\section{Tachyonic gluon masses}

\subsection{$\langle A^{A}_{\mu}A^B_{\nu}\rangle^{-1}$}

     We consider the diagrams in Fig.~B1(a) in the limit $p\to 0$, where $p$ is the external momentum.  
The ghost propagators in Eq.(\ref{301}) and the interaction   
\[
\sum_{A=3,8}\left[-gA^{A\mu}\sum_{\alpha=1}^3 \epsilon_{\alpha}^A\left\{ (\ptl_{\mu}\bar{C}^{\alpha)})C^{-\alpha}
- (\ptl_{\mu}\bar{C}^{-\alpha})C^{\alpha} \right\}\right]
\]
in Eq.(\ref{302}) gives the integral  
\begin{equation}
\sum_{\alpha=1}^3 g^2\epsilon^A_{\alpha} \epsilon^B_{\alpha} i \int \frac{d^4k}{(2\pi)^4}
\left\{\frac{k_{\mu}k_{\nu}(-k^2+i\epsilon^3_{\alpha}v)^2}{[k^4+(\epsilon^3_{\alpha}v)^2]^2}
+(v \to -v)\right\}.  \lb{b01}
\end{equation}
Performing the Wick rotation and neglecting $v$-independent terms, we obtain the formula 
\begin{align}
& i \int \frac{d^4k}{(2\pi)^4}
\left\{\frac{k_{\mu}(-k^2+i\eta v)k_{\nu}(-k^2+i\xi v)}{[k^4+(\eta v)^2][k^4+(\xi v)^2]} +(v \to -v)\right\} \nonumber \\
& =2i \int \frac{d^4k}{(2\pi)^4}
\frac{k_{\mu}k_{\nu}(k^4 - \eta \xi v^2)}{[k^4+(\eta v)^2][k^4+(\xi v)^2]} 
= -\frac{v}{64\pi}g_{\mu\nu}\frac{\eta^2+\xi^2 +\eta \xi +|\eta \xi|}{|\eta| +|\xi|}.   \lb{b02}
\end{align}
If we apply this formula, we find Eq.(\ref{b01}) becomes 
\begin{equation*}
-\frac{g^2v}{32\pi}g_{\mu\nu}\epsilon^A_{\alpha}\epsilon^B_{\alpha}|\epsilon^3_{\alpha}|.   
\end{equation*}  
Using the values of $\epsilon^A_{\alpha}$ in Eq. (\ref{202}), we obtain 
\begin{equation}
 -\frac{5}{2}g_{\mu\nu}m^2\ (A=B=3),\quad -\frac{3}{2}g_{\mu\nu}m^2 \ (A=B=8), \quad m^2=\frac{g^2v}{64\pi}.  \lb{b03}
\end{equation}

\begin{figure}
\begin{center}
\includegraphics{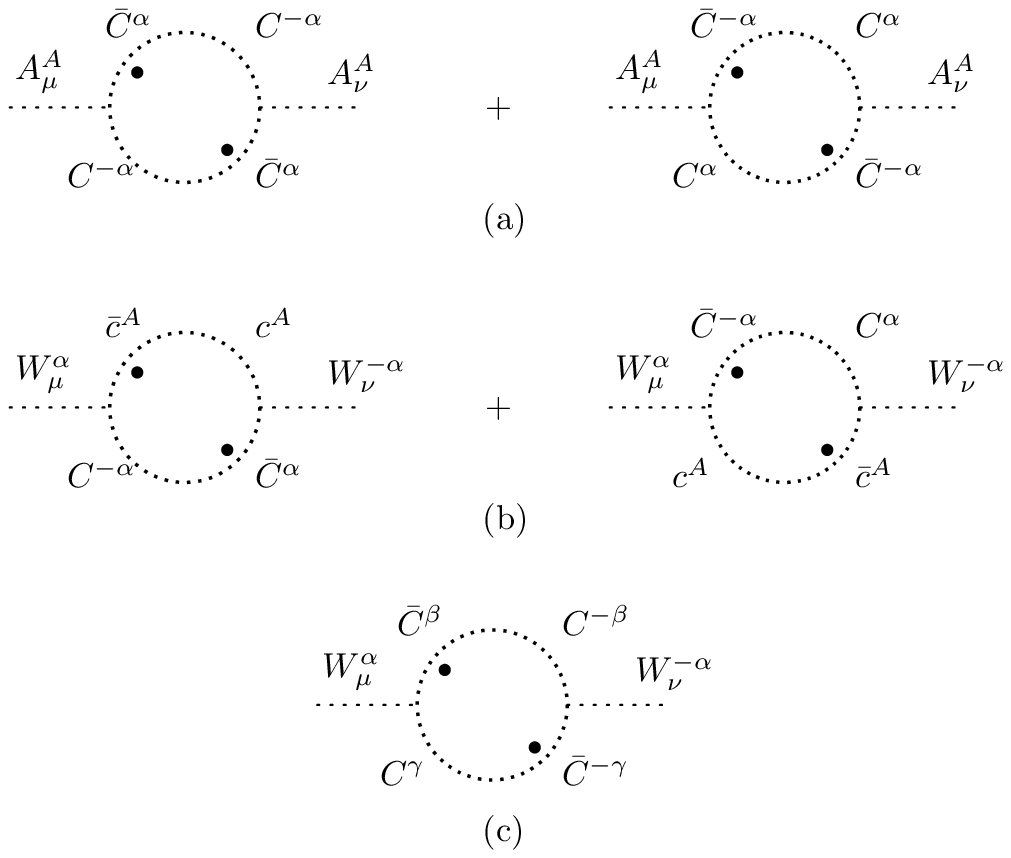}
\caption{The diagrams that contribute to the tachyonic gluon masses.  }
\label{fig5}
\end{center}
\end{figure}

\subsection{$\langle W^{\alpha}_{\mu}W^{-\alpha}_{\nu}\rangle^{-1}$}

    The diagrams in Fig.~B1(b), which come from the interaction 
\begin{equation*}
 \sum_{A=3,8}\left[
g(\ptl_{\mu}\bar{c}^A)\sum_{\alpha=1}^3\epsilon_{\alpha}^A(W^{\alpha\mu}C^{-\alpha}-W^{-\alpha\mu}C^{\alpha}) 
-g\sum_{\alpha=1}^3\epsilon_{\alpha}^A\left\{W^{\alpha\mu}(\ptl_{\mu}\bar{C}^{-\alpha})-
W^{-\alpha\mu}(\ptl_{\mu}\bar{C}^{\alpha})\right\}c^A \right],
\end{equation*}
give the integral   
\begin{equation*}
\sum_{A=3,8} g^2(\epsilon^A_{\alpha})^2 i \int \frac{d^4k}{(2\pi)^4}
\left\{\frac{k_{\mu}k_{\nu}(-k^2)(-k^2+i\epsilon^3_{\alpha}v)}{k^4[k^4+(\epsilon^3_{\alpha}v)^2]}
+(v \to -v)\right\}=
-\sum_{A=3,8}\frac{g^2v}{64\pi}g_{\mu\nu}(\epsilon^A_{\alpha})^2|\epsilon^3_{\alpha}|.   
\end{equation*}
Using the values of $\epsilon^A_{\alpha}$ in Eq.(\ref{202}),  we obtain 
\begin{equation}
   -g_{\mu\nu}m^2 \ (\alpha=1),\quad -\frac{1}{2}g_{\mu\nu}m^2 \ (\alpha=2, 3).  \lb{b04}
\end{equation}

     In the same way,  the interaction 
\begin{equation*}
\sum_{(\alpha,\beta,\gamma)}\mathrm{sgn}(\gamma)\frac{g}{\sqrt{2}}\epsilon_{\alpha\beta\gamma}
(\ptl_{\mu}\bar{C}^{\alpha})C^{\beta}W^{\gamma \mu}
\end{equation*}
in Eq.(\ref{302}) produces the diagram in Fig.~B1(c).  Applying the formula (\ref{b02}), this diagram gives    
\begin{align}
 &  \frac{g^2}{2} i \int \frac{d^4k}{(2\pi)^4}
\left\{\frac{k_{\mu}k_{\nu}(-k^2+i\epsilon^3_{\beta}v)(-k^2-i\epsilon^3_{\gamma}v)}{[k^4+(\epsilon^3_{\beta}v)^2][k^4+(\epsilon^3_{\gamma}v)^2]}
+(v \to -v)\right\} \nonumber \\
&=
-\frac{g^2v}{64\pi}g_{\mu\nu}\frac{(\epsilon^3_{\beta})^2+(\epsilon^3_{\gamma})^2-\epsilon^3_{\beta}\epsilon^3_{\gamma}+|\epsilon^3_{\beta}\epsilon^3_{\gamma}|}
{|\epsilon^3_{\beta}|+|\epsilon^3_{\gamma}|} \quad (\beta<\gamma,\  \alpha\neq \beta, \gamma).  \lb{b05}
\end{align}
From the values of $\epsilon^A_{\alpha}$ in Eq.(\ref{202}), we find Eq.(\ref{b05}) becomes   
\begin{equation}
   -\frac{1}{4}g_{\mu\nu}m^2 \ (\alpha=1),\quad -\frac{3}{4}g_{\mu\nu}m^2 \ (\alpha=2, 3).  \lb{b06}
\end{equation} 

    Thus, by summing up Eqs.(\ref{b04}) and (\ref{b06}), we obtain 
\begin{equation}
   -\frac{5}{4}g_{\mu\nu}m^2 \ (\alpha=1,2,3).  \lb{b07}
\end{equation}

\section{Example of the electric potential and its dual potential}

     In this appendix, we present an example of the massless electric potential 
$\bt^A_{\mu}$ and its dual potential $\mcb^A_{\mu}$ for a color electric charge $Q^A$.   
The color electric current is 
\[    j^{A\mu}= Q^A \delta(x)\delta(y)\delta(z)g^{\mu 0}, \] 
and the electric potential 
\begin{equation}
      \bt^{A\mu}=\frac{Q^A}{4\pi}\frac{1}{r}g^{\mu 0}, \quad r=\sqrt{x^2+y^2+z^2} \lb{c01}
\end{equation}
satisfies the equation of motion 
\[       \ptl_{\mu}(\ptl \wedge \bt^A)^{\mu\nu}-j^{A\nu}=0. 
 \]

     The dual electric potential that corresponds to (\ref{c01}) is 
\begin{equation}
      \mcb^{A\mu}=\frac{Q^A}{4\pi}\frac{z-r}{r\rho^2}(0,-y,x,0), \quad \rho=\sqrt{x^2+y^2}.  \lb{c02}
\end{equation}
This field fulfills the equation of motion 
\begin{equation*}
      \ptl_{\mu}(\ptl \wedge \mcb^A)^{\mu\nu} +
\epsilon^{\nu\alpha\mu\beta}\frac{n_{\alpha}\ptl_{\mu}}{n\cdot \ptl}j^A_{\beta}=0,     
\end{equation*}
and gives the color electric field  
\begin{equation}
  E^{Ai}=-\epsilon^{i0jk}\ptl_{j}\mcb^A_{k} = \frac{Q^A}{4\pi}\frac{x^i}{r^3} +\delta^i_3 Q^A\delta(x)\delta(y)\theta(-z),   \lb{c03}
\end{equation}
where $\theta(z)$ is the unit step function, and $(\ptl_x^2+\ptl_y^2)\ln \rho =2\pi \delta(x)\delta(y)$ has been used.  

     From Eq.(\ref{c01}), we get 
\[  (\ptl\wedge \bt^A)^{i0}= \frac{Q^A}{4\pi}\frac{x^i}{r^3}.  \]
The string part in Eq.(\ref{c03}) comes from $\Lambda_e^{A\mu\nu}$ in Eq.(\ref{405}).  
To choose the electric string in the negative $z$-direction, we use 
\begin{equation}
   \frac{1}{\ptl_z}\delta(z)=-\theta(-z).  \lb{c04}
\end{equation}
Then  Eq.(\ref{c01}) gives 
\[  (\Lambda_e^A)^{i0}=- \delta^i_3 \frac{1}{\ptl_z}\ptl_j(\ptl\wedge \bt^A)^{j0} =  \delta^i_3Q^A\delta(x)\delta(y)\theta(-z), 
\]
where $\nabla^2(1/r)=-4\pi \delta(\bve{r})$ has been used.  The sum $(\ptl\wedge \bt^A)^{i0}+ (\Lambda_e^A)^{i0}$ 
reproduces Eq.(\ref{c03}).

\section{The potentials $V_{Y}^A(r)$ and $V_{L}^A(r)$}

\subsection{ $V_{Y}^A(r)$}

     By subtracting $r$-independent terms, which contain ultraviolet divergence,  $V_{Y}^A$ in Eq.(\ref{407}) becomes     
\[ \int \frac{d^3q}{(2\pi)^3}\frac{Q^A_aQ^A_b}{q^2+m_A^2}e^{i \bve{q}\cdot \bve{r}}.  \] 
If the mass $m_A$ disappears above some scale $\Lambda_c$, this potential can be written as 
\begin{equation*}
     \int_0^{\Lambda_c} dq W(\bve{q},m,r) + \int_{\Lambda_c}^{\infty} dq W(\bve{q},0,r)= \int_{0}^{\infty} dq W(\bve{q},0,r)
    + \int_0^{\Lambda_c} dq \left\{W(\bve{q},m,r) -W(\bve{q},0,r)\right\}.  
\end{equation*}
The first term on the right hand side gives the Coulomb potential, which contributes mainly in the small $r$ region. 
When $r$ becomes large, the second term weakens the effect of the first term.  After performing the integration, we 
obtain \cite{hs21}
\begin{equation}
  V_Y^A(r)=Q^A_aQ^A_b\left( \frac{1}{4\pi r}-\frac{m_A^2}{2\pi^2}\int_0^{\Lambda_c} dq \frac{\sin qr}{qr}\frac{1}{q^2+m_A^2}\right).  
\lb{d01}
\end{equation}
We note this potential satisfies 
\begin{equation} 
  \lim_{\Lambda_c \to \infty}V_Y^A(r) = \frac{Q^A_aQ^A_b}{4\pi}\frac{e^{-m^Ar}}{r}.   \lb{d02}
\end{equation}
In the usual approach \cite{suz, mts, sst, hs17}, the cut-off is not taken into account, and $V_Y^A(r)$ 
becomes the Yukawa potential (\ref{d02}).

\subsection{ $V_{L}^A(\bve{r})$}

     When $m^A=0$, the potential $V_L^A(\bve{r})$ in Eq.(\ref{408}) vanishes.  So, different from $V_Y^A(r)$, the momentum region 
$q=|\bve{q}| \leq \Lambda_c$ contributes to  $V_L^A(r)$.  
Let us write $\bve{r}=(r_n, \bve{r}_t)$, and choose $\bve{n}$ as $r_n=\bve{r}\cdot \bve{n} \geq 0$.  The vector $\bve{r}_t$ satisfies 
$\bve{r}_t\cdot \bve{n}=0$, and $r_t=|\bve{r}_t|$.  
Similarly, we write $\bve{q}=(q_n, \bve{q}_t)$, and use the spherical coordinates  
\begin{equation}
  q_n=q\cos \theta,\ q_{t1}=q\sin\theta \cos \vp, \  q_{t2}=q\sin\theta \sin \vp,\ (q<\Lambda_c,\ 0\leq \theta\leq \pi,\ 
0\leq \vp <2\pi),  \lb{d03}
\end{equation}
where $q_{t1}$ is chosen to satisfy $\bve{q}_t\cdot \bve{r}_t=q\sin\theta \cos\vp r_t$.

\begin{figure}
\begin{center}
\includegraphics{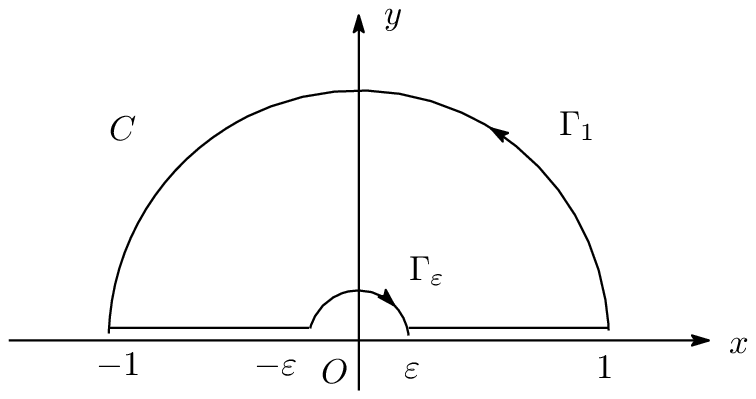}
\caption{The path $C$ on the complex plane.  }
\label{fig6}
\end{center}
\end{figure}

     Now we consider the integral 
\begin{equation}
   \int \frac{d^3q}{(2\pi)^3} \frac{e^{i\bve{q}\cdot \bve{r}}}{q_n^2(q^2+m_A^2)}  \lb{d04} 
\end{equation}
in $V_L^A$.  It becomes  
\begin{equation}
   \int_0^{\Lambda_c} \frac{dq}{(2\pi)^3} \int_0^{\pi} d\theta \sin\theta \int_0^{2\pi} d\vp 
\frac{e^{i q r_n\cos\theta}e^{iqr_t \sin\theta \cos\vp}}{\cos^2\theta (q^2+m_A^2)}.  \lb{d05}
\end{equation}
By changing the variable $\theta$ to $u=\cos \theta$, we get 
\begin{equation*}
   \int_0^{\pi} d\theta \sin\theta \frac{e^{i q r_n \cos\theta}e^{iqr_t \sin\theta \cos\vp}}{\cos^2\theta }
=\int_{-1}^1 du \frac{e^{i qr_n u}e^{iqr_t \sqrt{1-u^2} \cos\vp}}{u^2},   
\end{equation*}
which diverges at $u=0$.  If we choose the path $C$ in Fig.~D1, the integral 
\begin{equation*}
    \int_C dz \frac{e^{i z qr_n}e^{i q r_t\sqrt{1-z^2}\cos \vp}}{z^2}=0,   
\end{equation*}
leads to  
\begin{align}
   & \mathcal{P}\int_{-1}^{1} du \frac{e^{i qr_n u }e^{i q r_t\sqrt{1-u^2}\cos\vp}}{u^2} = I_{\Gamma_{\ve}} +  I_{\Gamma_{1}}, \lb{d06}
 \\
  &  I_{\Gamma_{\ve}}=-\int_{\Gamma_{\ve}} dz \frac{e^{i q r_n z}e^{i q r_t\sqrt{1-z^2}\cos\vp}}{z^2},\quad  
  I_{\Gamma_{1}} =- \int_{\Gamma_{1}} dz \frac{e^{i q r_n z}e^{i q r_t\sqrt{1-z^2}\cos\vp}}{z^2},  \nonumber  
\end{align}
where $\mathcal{P}$ means the Cauchy principal value.  
To calculate $I_{\Gamma_{\ve}}$, we use the variable $z=\ve e^{i\phi}$, and take the limit $\ve \to +0$.  Then it becomes 
\begin{equation}
     I_{\Gamma_{\ve}}= \lim_{\ve\to +0} \left\{ \frac{2}{\ve}-\pi q r_n +O(\ve)\right\}e^{i q r_t\cos\vp}.  \lb{d07}
\end{equation}  

    Similarly, by setting $z=e^{i\phi}$ in $ I_{\Gamma_{1}}$, we find 
\begin{equation}
    I_{\Gamma_1}= -i\int_0^{\pi} d\phi e^{-i\phi}e^{iq r_n e^{i\phi}}e^{iqr_t \cos \vp \sqrt{1-e^{2i\phi}}}.  \lb{d08} 
\end{equation}
We note Eq.(\ref{d08}) satisfies 
\begin{equation}
   \left| I_{\Gamma_{1}}\right|\leq \int_0^{\pi} d\phi e^{-q r_n\sin\phi}
e^{-q r_t\cos\vp\sqrt{2\sin \phi}\left\{\sin (2\phi-\pi)/4\right\}}
\leq \pi e^{qr_t/2}.  \lb{d09}
\end{equation}

     If we substitute Eqs. (\ref{d06}), (\ref{d07}) and (\ref{d08}) into Eq.(\ref{d05}), we find  
\begin{align}
   \int \frac{d^3q}{(2\pi)^3} \frac{e^{i\bve{q}\cdot \bve{r}}}{q_n^2(q^2+m_A^2)} &=
  \frac{1}{2\pi^2 \ve}H(m_A,\Lambda_c, r_t)-\frac{1}{4\pi}K_0(m_A r_t,\Lambda_c) r_n 
   + \mathcal{I}(m_A,\Lambda_c,r_n,r_t),  \lb{d10} \\
   \mathcal{I}(m_A,\Lambda_c,r_n,r_t)&=\int_0^{\Lambda_c}\frac{dq}{(2\pi)^3} \frac{1}{q^2+m_A^2}\int_0^{2\pi} d\vp \ I_{\Gamma_1},  
 \nonumber 
\end{align}
where,  
using the Bessel function $J_0(qr_t)$, $H(m_A, \Lambda_c, r_t)$ and  $K_0(m_Ar_t, \Lambda_c)$ are defined by  
\begin{align}
  H(m_A, \Lambda_c, r_t)&=  \int_0^{\Lambda_c} dq \frac{1}{q^2+m_A^2}J_0(qr_t), \quad
  K_0(m_Ar_t, \Lambda_c)=\int_0^{\Lambda_c} dq \frac{q}{q^2+m_A^2}J_0(qr_t), \lb{d11} \\
 J_0(qr_t)&=\frac{1}{2\pi}\int_{0}^{2\pi} d\vp e^{i qr_t \cos \vp}.  \nonumber
\end{align}
These functions satisfy  
\begin{align}
 H(m_A, \Lambda_c, 0)&=\frac{1}{m_A}\mathrm{tan}^{-1}\frac{\Lambda_c}{m_A}, 
\quad \lim_{r_t\to +0}K_0(m_Ar_t, \Lambda_c) = \frac{1}{2}\ln \left(\frac{\Lambda_c^2+m_A^2}{m_A^2}\right),  \lb{d12} \\
 K_0(m_Ar_t)&=\lim_{\Lambda_c \to \infty} K_0(m_Ar_t, \Lambda_c), \nonumber
\end{align}  
where $K_0(m_Ar_t)$ is the modified Bessel function.  
The term $\mathcal{I}$ in Eq.(\ref{d10}) has the properties   
\begin{align}
     \mathcal{I}(m_A,\Lambda_c, 0, 0)&=-\frac{1}{2\pi^2}\frac{1}{m_A}\tan^{-1}\frac{\Lambda_c}{m_A}, \lb{d13} \\
    \left|\mathcal{I}(m_A,\Lambda_c, r_n, r_t) \right|&
  \leq \frac{1}{8\pi^3}\int_0^{\Lambda_c} dq \frac{1}{q^2+m_A^2}\int_0^{2\pi} d\vp \left|I_{\Gamma_{1}}\right|
 \leq \frac{1}{4\pi} \int_0^{\Lambda_c} dq \frac{1}{q^2+m_A^2}e^{q r_t/2} \nonumber \\
 &
\to \ \frac{1}{4\pi m_A}\tan^{-1}\frac{\Lambda_c}{m_A} \quad (r_t \to 0).  \lb{d14}
\end{align}

     Thus, using Eqs.(\ref{d10}), (\ref{d12}) and (\ref{d13}), the potential $V_L^A$ in Eq.(\ref{408}) becomes 
\begin{align}
   V_L^A(\bve{r})=&V_{\mathrm{IR}}^A(r_t) - \frac{Q_a^AQ_b^Am_A^2}{4\pi}K_0(m_Ar_t, \Lambda_c) r_n \nonumber \\
 &+ m_A^2\left\{-\frac{(Q^A_a)^2+(Q^A_b)^2}{2} \frac{1}{2\pi^2m_A} \tan^{-1} \frac{\Lambda_c}{m_A}
+ Q_a^A Q_b^A\mathcal{I}(m_A,\Lambda_c,r_n,r_t) \right\},  \lb{d15}\\
  V_{\mathrm{IR}}^A(r_t) =&
\frac{m_A^2}{2\pi^2 \ve}\left\{\frac{(Q^A_a)^2+(Q^A_b)^2}{2} \frac{1}{m_A} \tan^{-1} \frac{\Lambda_c}{m_A}
+ Q_a^A Q_b^A H(m_A,\Lambda_c,r_t)\right\}.  \lb{d16}  
\end{align}

     We note the first term has the infrared divergence $1/\ve$, and the second term leads to the linear potential.  
When $r_t \to 0$, as Eq.(\ref{d14}) shows, the last term does not depend on $r_n$ so much.  
Therefore, in Sect. 4, we study the potential $V_L^A(r)$ based on the first and the second terms in Eq.(\ref{d15}).

     We make a comment.  Usually, the ultraviolet cut-off $\Lambda_c$ is introduced as $|q_t|<\Lambda_c$ 
 \cite{suz, mts, sst, hs19}.  The domain of integration is $|q_n|<\infty$ and $|q_t|<\Lambda_c$.  The infrared divergence and the linear 
potential come from the region with $|q_n| =\ve_n\ (\ve_n\ll 1)$.  In this article, as $m_A=0$ above $\Lambda_c$, the domain of 
integration is $q=|\bve{q}|<\Lambda_c$. The infrared divergence and the linear potential result from $\cos \theta=\ve \ (\ve \ll 1)$.  
Although the linear potential in these references coincides with that in this article, the coefficient 
of the infrared divergent term is different.  From $q_n=q\cos \theta$, we find $\ve_n$ is related to $\ve$ as $\ve_n=q\ve$.  
By using this relation, the infrared divergence in Ref.~\cite{hs19} becomes (\ref{d16}).    

\section{Type of the dual superconductivity}

     In the Ginzburg-Landau (GL) theory of the superconductivity, the space dependence of an order parameter $\Phi$ is considered 
(See, e.g., Ref.~\cite{nk}).  
To see the coherence length, the $x$-dependence is introduced as $\Phi(x)=\Phi f(x)$ with $f(0)=0$ and $\lim_{x\to \infty}f(x)=1$.  
From the phenomenological Lagrangian for $\Phi(x)$, the function $f(x)$ satisfies the equation 
\begin{equation}
     \xi^2\frac{d^2 f(x)}{dx^2}=-\left[1-\left\{f(x)\right\}^2\right]f(x).  \label{e01}
\end{equation}
The solution is $\displaystyle f(x)=\tanh \frac{x}{\sqrt{2}\xi}$, and $\xi$ is the coherence length.  
The penetration depth $\lambda$ is determined by the mass of the magnetic field, and the parameter 
$\kappa=\lambda/\xi$ is called the GL parameter.  
When $\kappa<1/\sqrt{2}$ ($\kappa>1/\sqrt{2})$, the superconductor is called type-I (type-II).  

     In the following subsections, under some assumptions, we consider the coherence length and the GL parameter 
in the present model.   

\subsection{SU(2) case}

     First, we consider the SU(2) case.  In Refs.~\cite{hs17, hs19}, we showed that the tachyonic mass term for 
the off-diagonal component $A_{\mu}^{\pm}=(A_{\mu}^1 \pm A_{\mu}^2)/\sqrt{2}$ is $-m^2 A_{\mu}^+A^{-\mu}$, and 
the interaction in $-F_{\mu\nu}^2/4$ contains the term $-g^2(A_{\mu}^+A^{-\mu})^2/2$.  From these terms, we obtain 
the gauge field condensate $\mathcal{G}=\langle A_{\mu}^+ A^{-\mu}\rangle^{(0)}=-m^2/g^2$ at the one-loop level.  
This condensate makes the classical U(1) field $b_{\mu}$ massive.  As its mass term becomes $m^2b_{\mu}b^{\mu}$, the 
penetration depth of $b_{\mu}$ is $\lambda=1/\sqrt{2}m$.  

     Now we consider the spatial behavior of the condensate $\mathcal{G}$.  
Since $\mathcal{G}$ has the mass dimension 2, we assume its $x$-dependence is expressed by 
$\mathcal{G}(x)=\left\{\sqrt{\mathcal{G}}f(x)\right\}^2$ with $f(0)=0$ and $\lim_{x\to \infty}f(x)=1$.  
As $\mathcal{G}(x)$ depends on $x$, we introduce the kinetic energy in the form of $\left\{ \sqrt{\mathcal{G}}f'(x)\right\}^2$.  
Thus, using this kinetic term, the above tachyonic mass term and the interaction, 
we assume the following phenomenological Lagrangian 
for $\mathcal{G}(x)$:  
\begin{equation*}
  \mathcal{L}_{2\mathrm{ph}}=\eta^2\mathcal{G}\left\{\frac{d f(x)}{dx}\right\}^2 
  -m^2\mathcal{G}\left\{f(x)\right\}^2-\frac{g^2}{2}\left[\mathcal{G}\left\{f(x)\right\}^2\right]^2, 
\end{equation*}
where $\eta$ is a parameter to adjust the effect of the assumed kinetic term.  This Lagrangian leads to the equation 
\begin{equation*}
     \eta^2 \frac{d^2 f(x)}{dx^2}=-m^2f(x)-g^2\mathcal{G}\left\{f(x)\right\}^3=-m^2\left[1-\left\{f(x)\right\}^2\right]f(x), 
\end{equation*}
where $\mathcal{G}=-m^2/g^2$ has been used.  This equation implies $\xi=\eta/m$.  
From $\lambda=1/\sqrt{2}m$ and $\xi=\eta/m$, we find $\kappa=1/\sqrt{2}\eta$.  If $\eta\simeq 1$, it implies 
the border between type-I and type-II. 

\subsection{SU(3) case}

     As in the SU(2) case, we assume the $x$-dependent order parameters 
$\mathcal{G}^{\alpha}(x)=\left\{\sqrt{\mathcal{G}^{\alpha}}f_{\alpha}(x)\right\}^2$ $(\alpha=1,2,3)$ and 
$G^A(x)=\left\{\sqrt{G^A}\phi_A(x)\right\}^2$ $(A=3,8)$.  Using the tachyonic mass terms 
(\ref{304}), the interaction terms (\ref{305}) and the assumed kinetic terms with the same parameter $\eta$, we consider 
the phenomenological Lagrangian 
\begin{align*}
\mathcal{L}_{3\mathrm{ph}} =& \sum_{\alpha=1}^3 \mathcal{G}^{\alpha}\left\{\eta^2\left(\frac{d f_{\alpha}}{dx}\right)^2
 - \frac{5}{4}m^2\left(f_{\alpha}\right)^2\right] 
 +\sum_{A=3,8} G^{A}\left\{\eta^2\left(\frac{d \phi_{A}}{dx}\right)^2 
  - \frac{m_A^2}{2}\left(\phi_{A}\right)^2 \right\} \\
 &-\frac{g^2}{2}\sum_{\alpha=1}^3 (\mathcal{G}^{\alpha})^2\left\{\left(f_{\alpha}\right)^2\right\}^2 
 -\frac{g^2}{4}\sum_{\alpha\neq \beta} \mathcal{G}^{\alpha}\mathcal{G}^{\beta}\left(f_{\alpha}\right)^2\left(f_{\beta}\right)^2\\
 &-g^2\left\{G^3\mathcal{G}^1(\phi_3)^2(f_1)^2
 +\frac{1}{4}\left\{G^3(\phi_3)^2+3G^8(\phi_8)^2\right\}\sum_{\alpha=2}^3\mathcal{G}^{\alpha}\left(f_{\alpha}\right)^2\right\}.  
\end{align*}

     From $\mathcal{L}_{3\mathrm{ph}}$, we obtain the equations for $f_2(x)$ and $\phi_8(x)$ given by 
\begin{align}
\eta^2\frac{d^2 f_2}{dx^2}= &-\frac{5m^2}{4}f_2 \nonumber \\
& -g^2\left[ \mathcal{G}^2(f_2)^2
+\frac{1}{2}\left\{\mathcal{G}^1(f_1)^2 +\mathcal{G}^3(f_3)^2\right\}
+\frac{1}{4}\left\{G^3(\phi_3)^2+3G^8(\phi_8)^2\right\}\right]f_2, \label{e02}\\
\eta^2\frac{d^2 \phi_8}{dx^2}= &-\frac{3m^2}{4}\phi_8 -\frac{3g^2}{4}\left\{ \mathcal{G}^2(f_2)^2 +\mathcal{G}^3(f_3)^2\right\}\phi_8.  
\label{e03} 
\end{align}
If we assume the relation $f_{\alpha}(x)\simeq \phi_A(x)$ $(\alpha=1,2,3,\ A=3,8)$, 
these equations become 
\begin{align}
 \eta^2\frac{d^2 f_2}{dx^2}\simeq &-\frac{5m^2}{4}\left\{1-(f_2)^2\right\}f_2, \label{e04} \\
 \eta^2\frac{d^2 \phi_8}{dx^2}\simeq &-\frac{3m^2}{4}\left\{1-(\phi_8)^2\right\}\phi_8 , \label{e05}
\end{align}
where Eqs.(\ref{306}) and (\ref{308}) have been used.  In the same way, 
we find $f_1, f_3$ and $\phi_3$ also satisfy Eq.(\ref{e04}).  
Therefore, comparing these equations with Eq.(\ref{e01}), we find 
\begin{equation}
  f_{\alpha}(x) \simeq \phi_3(x)\simeq \tanh \frac{x}{\sqrt{2}\xi_3},\ \xi_3=\frac{2\eta}{\sqrt{5}m},\quad
\phi_8(x)\simeq \tanh \frac{x}{\sqrt{2}\xi_8},\ \xi_8=\frac{2\eta}{\sqrt{3}m}.  \label{e06}
\end{equation}

     Eq.(\ref{e06}) shows that we have to modify Eqs.(\ref{e04}) and (\ref{e05}) to satisfy 
the relation $f_{\alpha}\simeq \phi_3\neq \phi_8$.  If we use this relation, Eqs.(\ref{e02}) and (\ref{e03}) become 
\begin{align}
 \eta^2\frac{d^2 f_2}{dx^2}\simeq &-\frac{5m^2}{4}\left\{1-(f_2)^2\right\}f_2-\frac{3g^2}{4}G^8\left\{(\phi_8)^2-(f_2)^2\right\}f_2, \label{e07} \\
 \eta^2\frac{d^2 \phi_8}{dx^2}\simeq &-\frac{3m^2}{4}\left\{1-(\phi_8)^2\right\}\phi_8 
 -\frac{3g^2}{2}\mathcal{G}^2\left\{(f_2)^2-(\phi_8)^2\right\}\phi_8, \label{e08}
\end{align}
where Eqs.(\ref{306}) and (\ref{308}) have been used again.  
Now we use Eq.(\ref{309}), and rewrite the second terms on the right hand side as 
\begin{align*}
 -\frac{3g^2}{4}G^8\left\{(\phi_8)^2-(f_2)^2\right\}f_2=-\frac{m^2}{16}\delta_2(x)\left\{1-(f_2)^2\right\}f_2,\quad 
 \delta_2(x)=\frac{(f_2)^2-(\phi_8)^2}{1-(f_2)^2},  \\
 -\frac{3g^2}{2}\mathcal{G}^2\left\{(f_2)^2-(\phi_8)^2\right\}\phi_8=\frac{3m^2}{4}\delta_8(x)\left\{1-(\phi_8)^2\right\}\phi_8,\quad 
 \delta_8(x)=\frac{(f_2)^2-(\phi_8)^2}{1-(\phi_8)^2}.
\end{align*}
Then Eqs.(\ref{e07}) and(\ref{e08}) become 
\begin{align}
 \eta^2\frac{d^2 f_2}{dx^2}\simeq &-\frac{5m^2}{4}\left\{1+\frac{\delta_2(x)}{20}\right\}\left\{1-(f_2)^2\right\}f_2, \label{e09} \\
 \eta^2\frac{d^2 \phi_8}{dx^2}\simeq &-\frac{3m^2}{4}\left\{1-\delta_8(x)\right\}\left\{1-(\phi_8)^2\right\}\phi_8. \label{e10}
\end{align}
We note, as $|f_ 2|<1, |\phi_8|<1$ and $|f_ 2|> |\phi_8|$, $\delta_2$ and $\delta_8$ 
satisfy $0<\delta_a <1$ $(a=2,8)$.

     Since $\delta_2$ and $\delta_8$ depend on $x$, it is difficult to solve Eqs.(\ref{e09}) and (\ref{e10}).  However, Eq.(\ref{e05}) becomes Eq.(\ref{e10}), if we replace $3m^2/4$ with $3m^2(1-\delta_2)/4$.  Therefore, it is expected that the coherence length 
$\xi_{\mathrm{max}}$ obtained from Eq.(\ref{e10}) is longer than $\xi_8=2\eta/\sqrt{3}m$.  
From the masses for the classical fields in Eq.(\ref{312}), 
the corresponding penetration depth is $\lambda_8=\sqrt{2}/{\sqrt{3}m}$.  
If we can use $\xi_{\mathrm{max}}$ and $\lambda_8$, the GL parameter becomes 
$\displaystyle \kappa=\lambda_8/\xi_{\mathrm{max}}
<\lambda_8/\xi_8=1/\sqrt{2}\eta$.  If $\eta\simeq 1$, we can expect type-I.

\end{document}